\UseRawInputEncoding
\documentclass[twocolumn,astrosymb]{aastex631}
\graphicspath{{./}{figures/}}

\usepackage{amsmath,amsfonts}
\usepackage{algorithmic}
\usepackage{algorithm}
\usepackage{array}
\usepackage{textcomp}
\usepackage{url}
\usepackage{verbatim}
\usepackage{graphicx}
\usepackage{gensymb}
\usepackage{hhline}
\usepackage{mathrsfs}
\usepackage{microtype}
\usepackage[caption=false]{subfig}
\usepackage{multirow}

\defcitealias{dahal22}{D22}
\defcitealias{nunez22spie}{N22}
\defcitealias{nunez23asc}{N23}
\usepackage{hyperref} 

\usepackage{booktabs}
\usepackage{footnote}

\newcommand{\unit}[2]{${#1}\,\mathrm{{#2}}$}

\newcommand{\Psat}{P_\mathrm{sat}}
\newcommand{\Tc}{T_\mathrm{c}}
\newcommand{\pW}{\,\mathrm{pW}}
\newcommand{\mK}{\,\mathrm{mK}}
\newcommand{\GHz}{\,\mathrm{GHz}}
\newcommand{\Tbath}{T_\mathrm{bath}}

\submitjournal{ApJS}

\begin{document}

\title{High-Efficiency and Low-Noise Detectors for the Upgraded CLASS 90~GHz Focal Plane}

\newcommand{\JHU}{The William H. Miller III Department of Physics and Astronomy, Johns Hopkins University, Baltimore, MD 21218, USA}
\newcommand{\NASA}{NASA Goddard Space Flight Center, Greenbelt, MD 20771, USA}
\newcommand{\Harvard}{Center for Astrophysics, Harvard \& Smithsonian, Cambridge, MA 02138, USA}
\newcommand{\Norway}{Institute of Theoretical Astrophysics, University of Oslo, Oslo, Norway}
\newcommand{\MIT}{MIT Kavli Institute, Massachusetts Institute of Technology, Cambridge, MA 02139, USA}
\newcommand{\Chicago}{Department of Astronomy and Astrophysics, University of Chicago,  Chicago, IL 60637, USA}
\newcommand{\Radiosky}{Radiosky, San Pedro de Atacama, Chile}
\newcommand{\UCSC}{Departamento de Ingenier\'{i}a El\'{e}ctrica, Universidad Cat\'{o}lica de la Sant\'{i}sima Concepci\'{o}n, Alonso de Ribera 2850, Concepci\'{o}n, Chile}
\newcommand{\UMBC}{Center for Space Sciences and Technology, University of Maryland, Baltimore County, Baltimore, MD 21250, USA}
\newcommand{\Villanova}{Department of Physics, Villanova University,  Villanova, PA 19085, USA}
\newcommand{\upenn}{Department of Physics and Astronomy, University of Pennsylvania, 209 South 33rd Street, Philadelphia, PA 19104, USA}

\author[0000-0002-5247-2523]{Carolina N\'u\~nez}
\affiliation{\JHU}
\author[0000-0002-8412-630X]{John~W. Appel}
\affiliation{\JHU}
\author[0000-0003-3853-8757]{Rahul Datta}
\affiliation{\JHU}
\affiliation{\Chicago}
\author[0000-0001-8839-7206]{Charles~L. Bennett}\affiliation{\JHU}
\author{Michael~K. Brewer}
\author[0000-0003-2682-7498]{Sarah~Marie Bruno}
\affiliation{\JHU}
\author[0000-0001-8468-9391]{Ricardo Bustos}
\affiliation{\UCSC}
\author[0000-0003-0016-0533]{David~T. Chuss}
\affiliation{\Villanova}
\author{Nick Costen}
\affiliation{\NASA}
\author[0000-0002-0552-3754]{Jullianna Denes~Couto}
\affiliation{\Radiosky}
\author[0000-0002-1708-5464]{Sumit Dahal}
\affiliation{\NASA}
\author{Kevin~L. Denis}
\affiliation{\NASA}
\author[0000-0001-6976-180X]{Joseph~R. Eimer}
\affiliation{\JHU}
\author[0000-0002-4782-3851]{Thomas Essinger-Hileman}
\affiliation{\NASA}
\author[0000-0001-7466-0317]{Jeffrey Iuliano}
\affiliation{\upenn}\affiliation{\JHU}
\author[0000-0002-4820-1122]{Yunyang Li}
\affiliation{\JHU}
\author[0000-0003-4496-6520]{Tobias~A. Marriage}
\affiliation{\JHU}
\author{Jennette Mateo}
\affiliation{\NASA}
\author[0000-0002-4436-4215]{Matthew~A. Petroff}
\affiliation{\Harvard}
\author[0000-0001-7458-6946]{Rui Shi}
\affiliation{\JHU}
\author[0000-0003-4189-0700]{Karwan Rostem}
\affiliation{\NASA}
\author[0000-0003-3487-2811]{Deniz~A.~N. Valle}
\affiliation{\Radiosky}
\author[0000-0002-5437-6121]{Duncan Watts}
\affiliation{\Norway}
\author[0000-0002-7567-4451]{Edward~J. Wollack}
\affiliation{\NASA}
\author[0000-0001-6924-9072]{Lingzhen Zeng}
\affiliation{\Harvard}

\begin{abstract}

We present the in-lab and on-sky performance for the upgraded 90~GHz focal plane of the Cosmology Large Angular Scale Surveyor (CLASS), which had four of its seven detector wafers updated during the austral winter of 2022.
The update aimed to improve the transition-edge-sensor (TES) stability and bias range and to realize the high optical efficiency of the sensor design. Modifications included revised circuit terminations, electrical contact between the TES superconductor and the normal metal providing the bulk of the bolometer heat capacity, and additional filtering on the TES bias lines. The upgrade was successful: 94\% of detectors are stable down to 15\% of the normal resistance, providing a wide overlapping range of bias voltages for all TESs on a wafer. The  median telescope efficiency improved from $0.42^{+0.15}_{-0.22}$ to $0.60^{+0.10}_{-0.32}$ (68\% quantiles).  For the four upgraded wafers alone, median telescope efficiency increased to $0.65^{+0.06}_{-0.06}$. Given our efficiency estimate for the receiver optics, this telescope efficiency implies a detector efficiency exceeding $0.90$.  The overall noise-equivalent temperature of the \unit{90}{GHz} focal plane improved from \unit{19}{\mu K\sqrt{s}} to \unit{9.7}{\mu K\sqrt{s}}.

\end{abstract}


\section{Introduction}
\label{sec:intro}

\begin{figure*}
\centering
   \subfloat[]{\includegraphics[width=.3\textwidth]{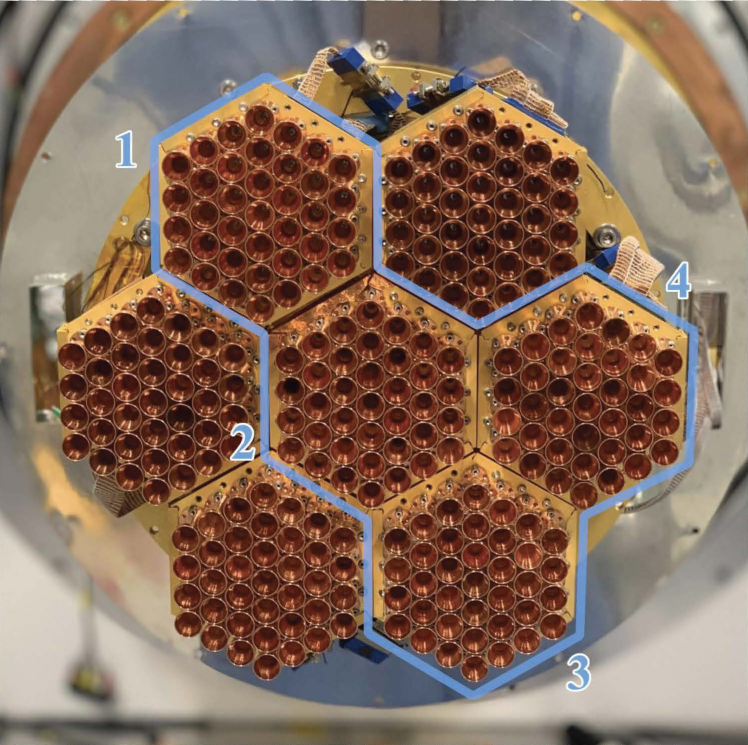}\label{subfig:W1_in_field}}
   \hfill
   \subfloat[]{ \includegraphics[width=.3\textwidth]{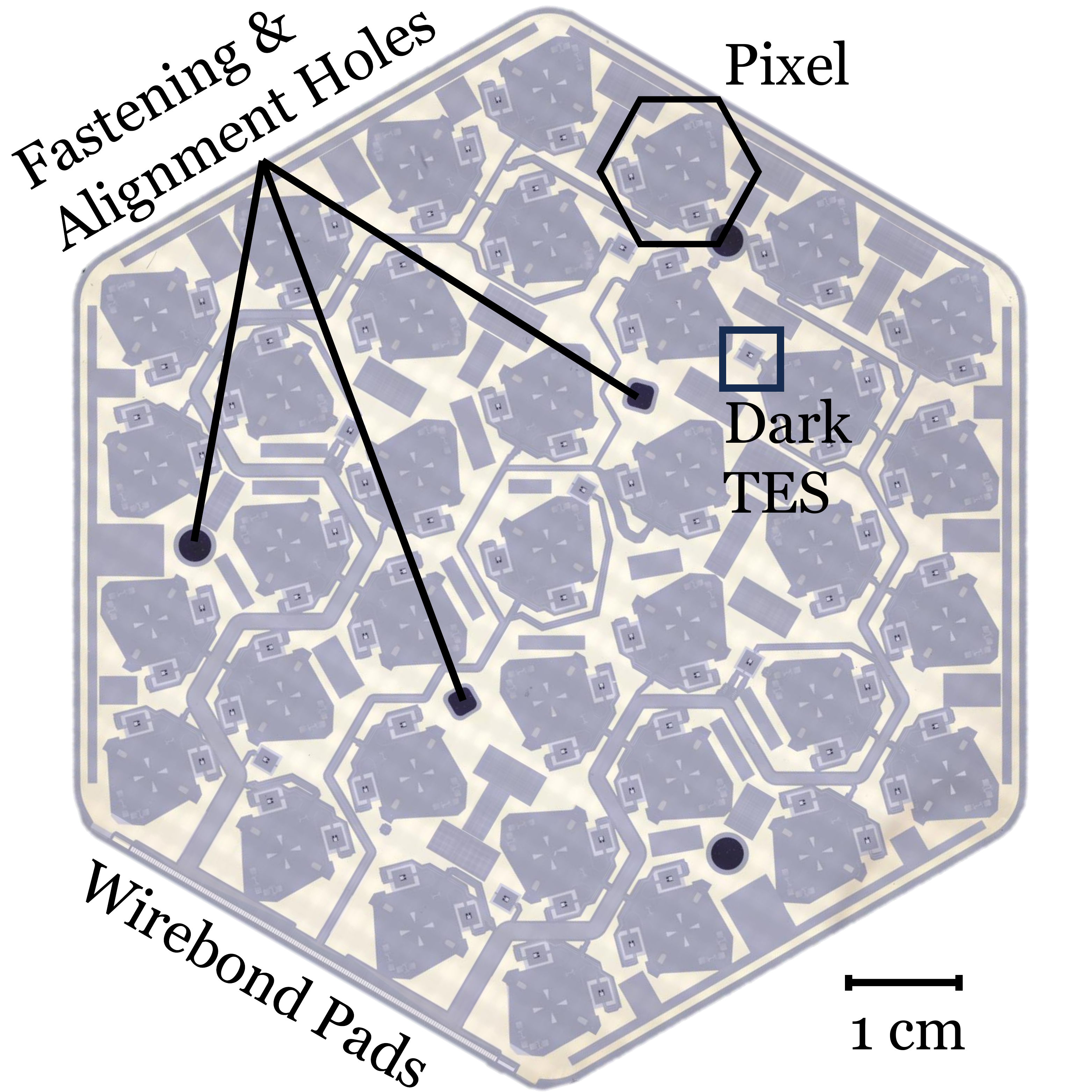}\label{subfig:w-band-detector-wafer_white}  } 
   \hfill
   \subfloat[]{\includegraphics[width=.3\textwidth]{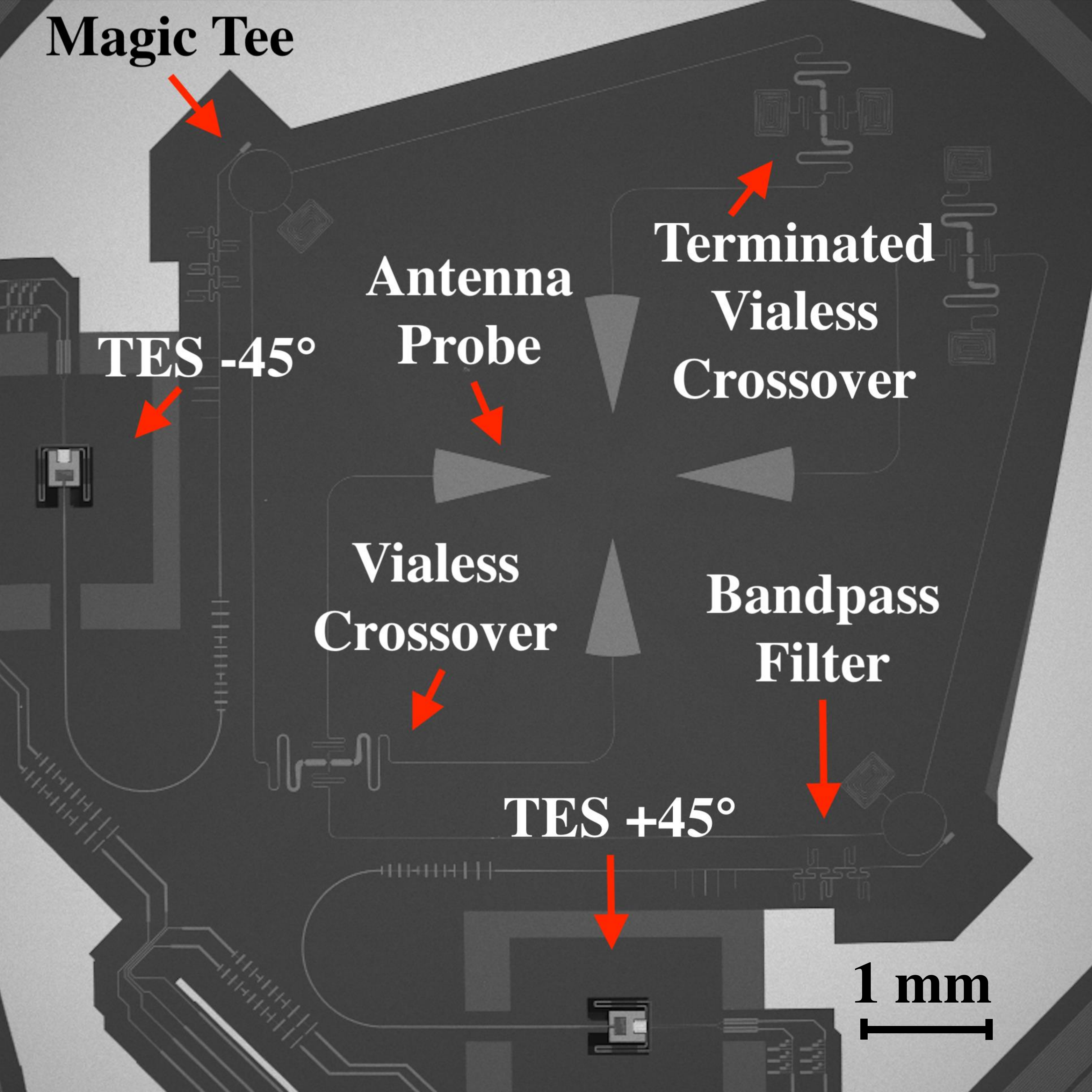} \label{subfig:revised_pixel}}\hfil
   \caption{\textbf{(a)} Photograph of the upgraded \unit{90}{GHz} focal plane, taken at the CLASS site.  The four new modules are outlined in blue and numbered with the convention used throughout the paper. Thirty-seven copper feedhorns per module couple detectors to the incoming radiation. \textbf{(b)} Within each module is an indium-bonded stack of silicon wafers into which is built an integrated array of thirty-seven dual-polarization detector pixels. \textbf{(c)} A single pixel shows updated components: the terminated vialess crossover, the magic tee, and the TES.  Images are reproduced from \citetalias{nunez22spie,nunez23asc}. }
\end{figure*}

Measurements of the cosmic microwave background (CMB) have provided the tightest constraints to date on the parameters of the $\Lambda$CDM cosmological model and its extensions \citep[e.g.,][]{bennett13,hinshaw13,planck18VI,BK21}. For the past twenty years, advances in the sensitivity of CMB measurements have been enabled in a large part by advances in cryogenic detector technology, with transition-edge-sensor (TES) bolometers playing a central role \cite[e.g.,][]{niemack08,kuo08,shirokoff10,arnold12,suzuki12,rostem12spie,appel14spie,posada16,hui16}.  Along these lines, this paper concerns improvements to cryogenic detector performance for the Cosmology Large Angular Scale Survey (CLASS) project. CLASS aims to measure CMB polarization on the largest angular scales ($\ell<30$) to constrain the optical depth of reionization and primordial gravitational radiation \citep{essinger-hileman14spie,watts15,harrington16spie,watts18}. This is one of the frontiers of the CMB field, with associated projects pursuing complementary strategies \citep{litebird22,piper14,lspe20,groundbird20,lspe20, quijote23, may2024}. Major upcoming international-scale ground-based surveys will constrain primordial gravitational waves and measure the growth of cosmic structure on smaller angular scales \citep{simons19whitepaper,stagefour22bmodeforecast}.

Located on Cerro Toco at \unit{5200}{m} in the Chilean Atacama Desert, the CLASS polarization-sensitive array of microwave telescopes operates in frequency bands centered approximately on 40, 90, 150, and \unit{220}{GHz} \citep{eimer12spie,appel14spie,harrington18spie,dahal18spie}. Among these, the \unit{90}{GHz} band is closest to the minimum in polarized Galactic emission \citep[][Figure 51]{planck15V}, and the nominal CLASS design has two dedicated \unit{90}{GHz} telescopes. However, the initial \unit{90}{GHz} detectors produced in 2018 had problematic variations in stable bias range and optical efficiency \citep[][hereafter \citetalias{dahal22}]{dahal22}. To improve detector performance we updated the detector design and replaced
part of the first 90 GHz telescope's detectors with the upgraded version in August 2022. Initial characterizations of the new detectors have been reported in \citet{nunez22spie,nunez23asc}, hereafter \citetalias{nunez22spie,nunez23asc}.

\begin{figure*}
    \centering
    \subfloat[]{%
    \includegraphics[height=2in]{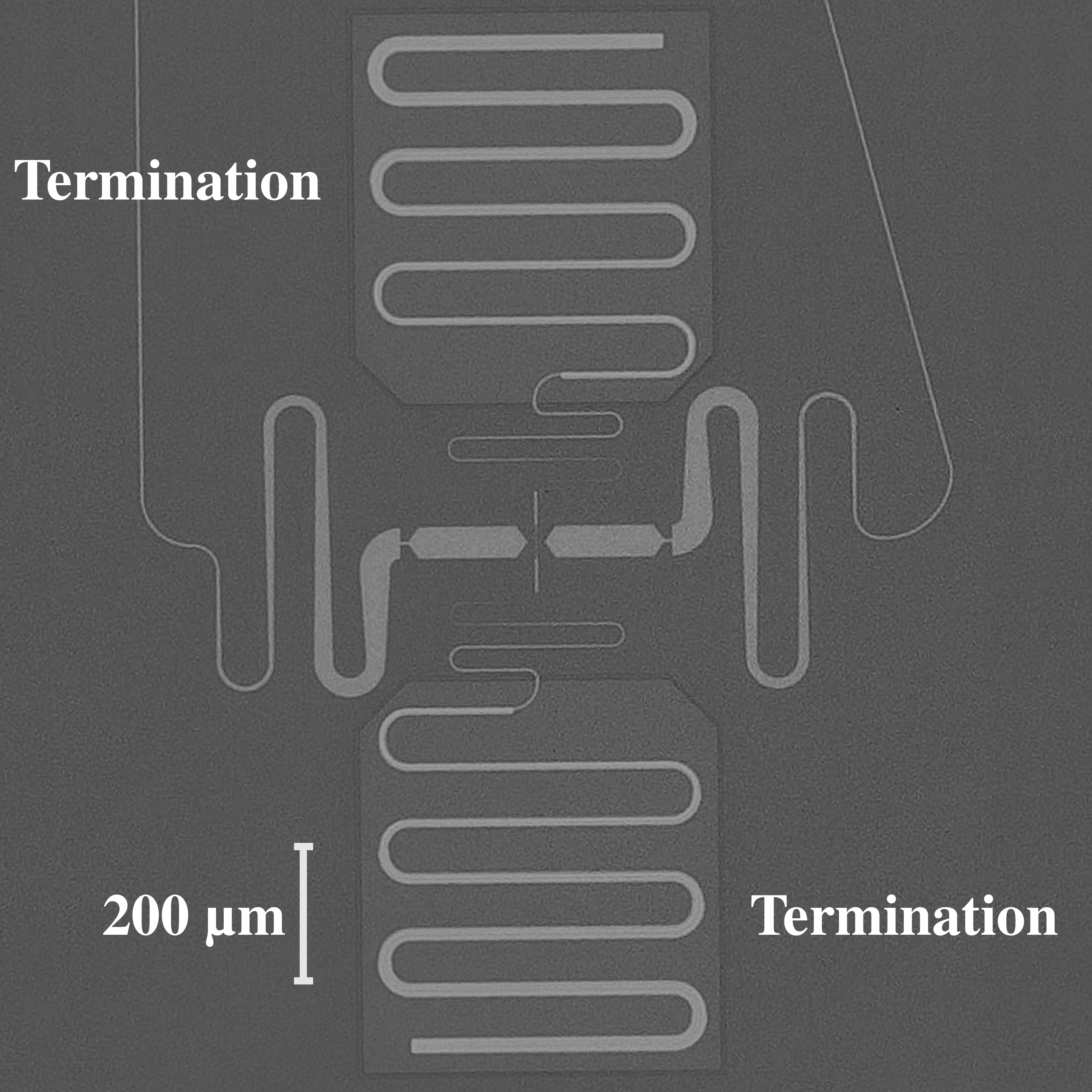}
    \label{subfig:crossover_OG}
    }\qquad
    \subfloat[]{%
    \includegraphics[height=2in]{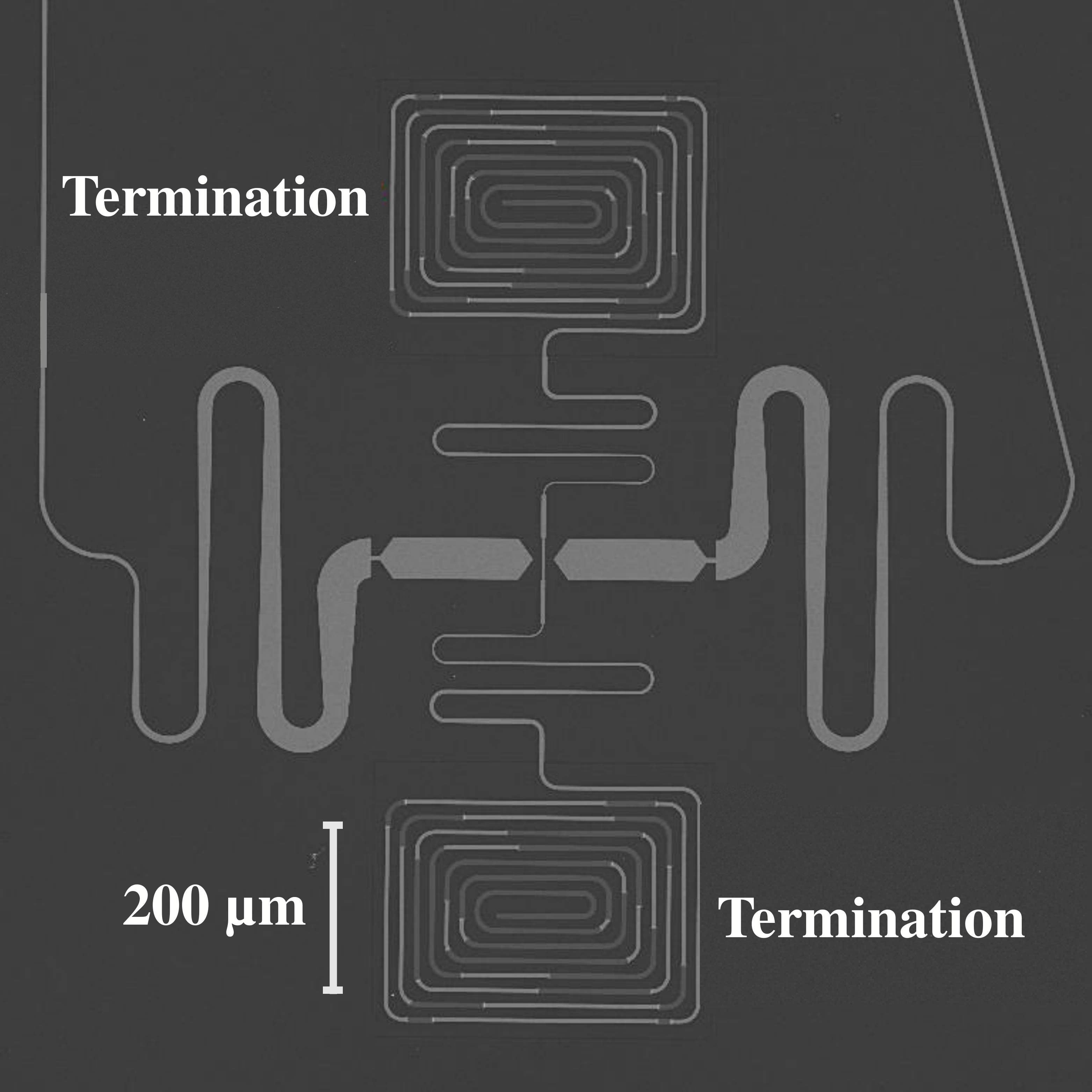}
    \label{subfig:crossover}
    }\qquad
    \subfloat[]{%
    \includegraphics[height=2in]{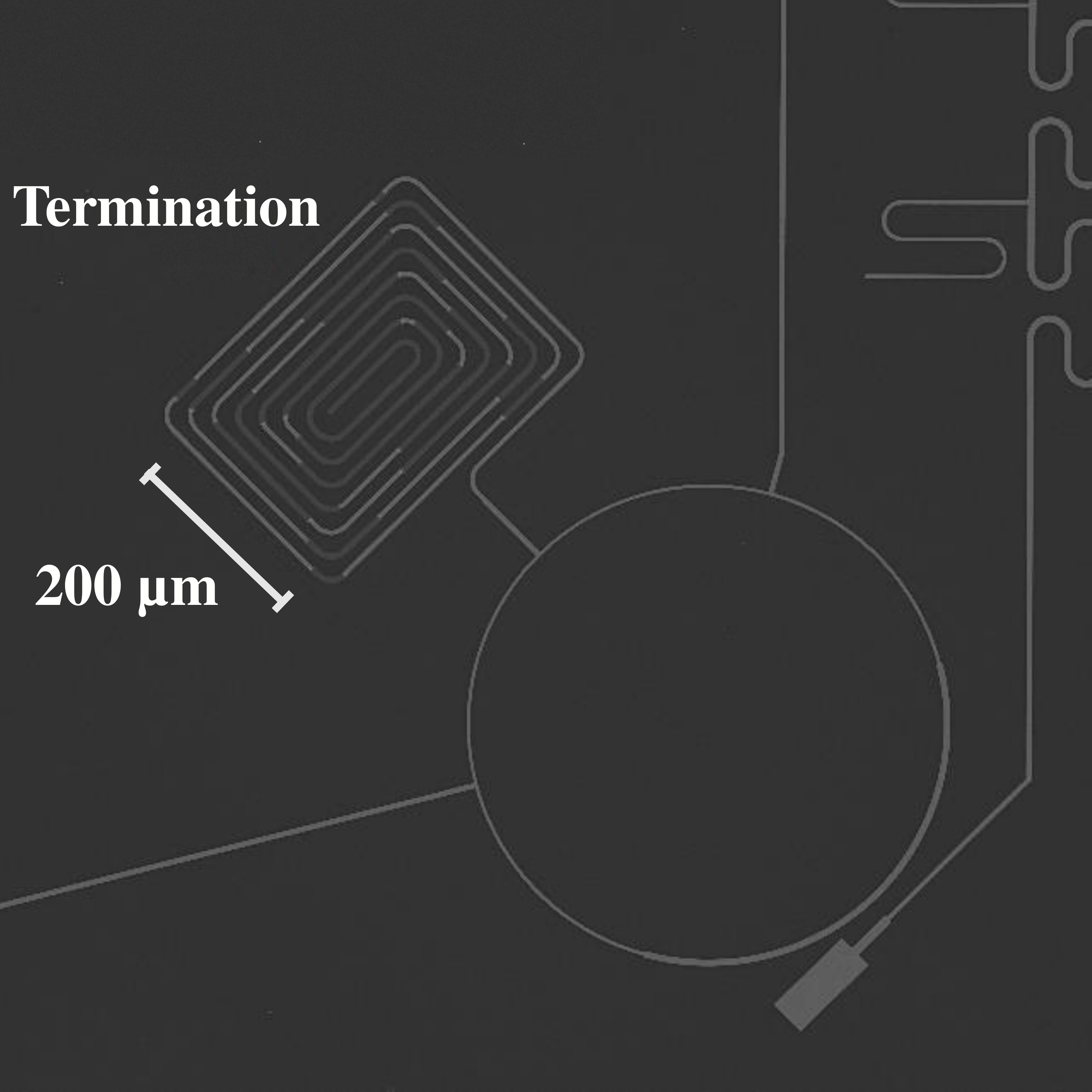}
    \label{subfig:magic_tee}
    }
    \caption{The terminated vialess crossover with the original \textbf{(a)} and updated \textbf{(b)} termination design together with the magic tee \textbf{(c)} with the updated design. In the original design, the superconducting Nb line is overlaid with a wider gold line for adiabatic absorption and impedance matching. The new termination was realized as a stepped impedance transformer from Nb to PdAu. (The light sections of the termination line are Nb while the dark sections are PdAu.) This design avoids proximitization, is robust against fabrication variability, and reduces the device footprint. Images with the revised design were adapted from \citetalias{nunez23asc}.}
     \label{fig:redesign}
\end{figure*}

In this paper, we present the full in-lab and on-sky characterization of the upgraded \unit{90}{GHz} focal plane.  We discuss the design of the upgraded detector wafers in Section~\ref{sec:design}, emphasizing updates. The types of data for detector characterization are described in Section~\ref{sec:data}.  Sections~\ref{sec:electrothermal}--\ref{sec:stability} cover the electrothermal TES parameters, bandpass, optical efficiency, noise performance, and stability. Finally, in Section~\ref{sec:conclusion} we summarize the results, while spatial variations in electrothermal properties and optical efficiency are covered in Appendix~\ref{app:Detector Property Spatial Variation}. An index of parameters used to describe the instrumentation is in Appendix~\ref{app:parameter index}.

\section{Upgraded Detectors} \label{sec:design}

Figure \ref{subfig:W1_in_field} shows a photograph of the four upgraded detector modules installed in the 90~GHz focal plane in July 2022. In total, the detector array holds seven modules, and those replaced were the least-well performing modules in the original array (W1). The most apparent features of the modules in the photo are the thirty-seven copper feedhorns that couple the incoming radiation to the detector pixels inside the module \citep{zeng10spie,dahal18spie}. Visible behind the feedhorns is the gold-plated aluminum-silicon alloy interface plate \citep{Ali-CE7}. Mounted to the back of this plate is an indium-bonded and gold-plated stack of silicon wafers that provides broadband optical coupling, out-of-band light rejection, and thermalization for the detector wafer shown in Figure 1b. Orthogonal polarized signals received via the feedhorn are diplexed by a symmetric waveguide orthomode transducer (OMT) and then subsequently filtered and detected on-chip. The detector wafer integrates thirty-seven dual-polarization detector pixels. The multi-layer detector stack is fastened and aligned with the interface and feedhorns through micromachined holes. In addition to the seventy-four TESs associated with detector pixels (one per polarization per pixel), several optically isolated ``dark'' TESs  are used to assess parasitic power loading. Each TES is voltage biased and read out through a pair of leads that connect to wirebond pads on one side of the wafer. Additional cryogenic module circuitry comprises shunt-resistor chips to provide voltage biases and time-division multiplexing chips to read the detector current \citep{reintsema03,nist_tdm_mux13b}. The detector modules are controlled and read out with warm Multi-Channel Electronics (MCE) \citep{Battistelli08}.

Figure~\ref{subfig:revised_pixel} shows the upgraded detector pixel, which inherits most of its design from the original W1 detectors. For a full description of the original CLASS 90~GHz module and detector design, see~\cite{Denis-fabrication, Chuss-development,  rostem16spie, dahal18spie}. Here we provide an overview and highlight design elements that changed in the upgraded detectors. Aside from terminations, the microwave circuitry on the detector wafer is made from superconducting niobium with a dielectric of single-crystal silicon derived from the \unit{5}{\mu m}-thick device layer of a silicon-on-insulator (SOI) wafer. A pair of symmetric antenna probes couple each orthogonal polarization of incoming light arriving through a circular waveguide that runs from the base of the feedhorn through the interface plate and the initial layers of the detector wafer stack. A waveguide quarter-wave backshort behind the detector wafer enhances the coupling \citep{Crowe-choke}. 
Transmission lines from the antenna probes transition from co-planar waveguide to microstrip transmission lines. 
The radiation from opposing probe antennas is then coherently combined at the magic tee, which transmits the difference and terminates the sum signal \citep{U-Yen-magicT}. The signal paths are symmetrized by passing through either a vialess crossover or an impedance-matched ``dummy'' termination~\citep{Uyen2009}.  The signal transmitted by the magic tee propagates through microstrip band-defining filters before terminating at the corresponding TES bolometer. 

\subsection{Revised Magic Tee and Crossover Terminations}
For the original 90 GHz detector pixels, absorber terminations for the teminated vialess crossovers and the magic tees were realized in a microstrip configuration by overlaying a superconducting Nb line with a wider resistive Au layer shaped to provide adiabatic absorption along the structure and impedance matching. An example of this termination design is shown in Figure \ref{subfig:crossover_OG}. Due to concerns that termination performance was degraded by proximitization of the Au \citep{proximity2, proximity1}, the terminations for the upgraded pixel were redesigned. Images of the crossover and magic tee with the new termination design are shown in Figures~\ref{subfig:crossover} and \ref{subfig:magic_tee}. The new design features a stepped-impedance transformer from superconducting Nb to normal PdAu. 
Palladium's larger density of states, reduced diffusivity, and strong Pauli paramagnetism combine to limit the proximitization of PdAu relative to that of Au. 
PdAu is a disordered alloy (RRR=1.1) with a relatively high electrical bulk resistivity, allowing a reduction in the termination footprint. Finally, the stepped-impedance design is more robust against dimensional variations and misalignments in fabrication than the original tapered design.

\begin{figure}
\centering
\includegraphics[width=.45\textwidth]{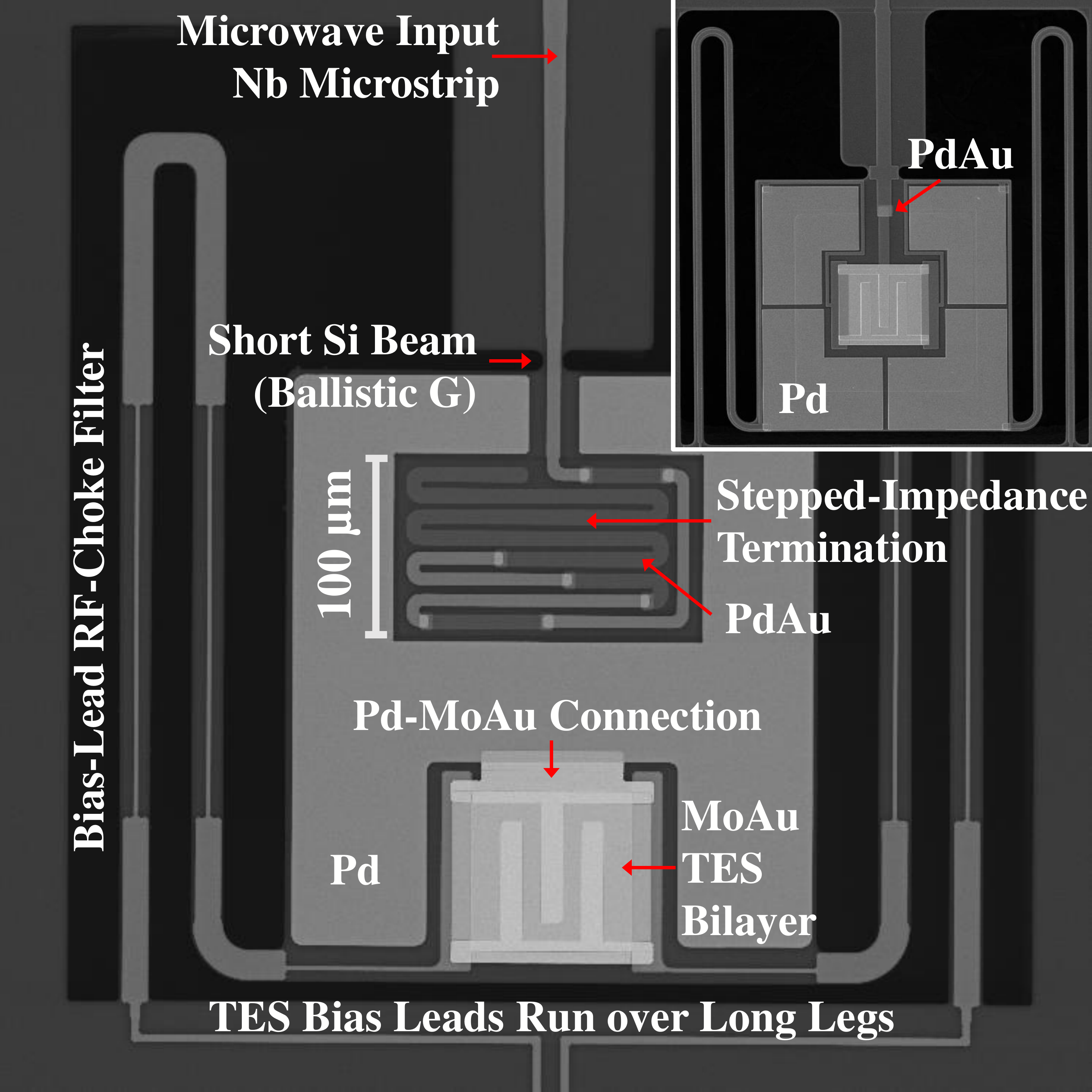} 
\caption{The $90\GHz$ TES Bolometer. (Main Image) The TES was revised to include (1) an Nb-PdAu stepped-impedance termination, (2) an single Pd film for thermal mass, (3) a direct normal-metal contact between the MoAu TES and the Pd, and (4) a revised bias-lead filter circuit implementation extending RF chokes onto the bias-lead legs. This image was reproduced from \citetalias{nunez23asc}. (Top-Right Inset) The original TES used a left-right symmetric, distributed quarter-wave backshort termination with PdAu absorber. This image was reproduced from \cite{dahal18spie}.}
\label{fig:TES_architecture}
\end{figure}

\subsection{Revised TES Bolometer Thermal Circuit}
\label{subsec:TES Thermal Circuit}

Figure \ref{fig:TES_architecture} shows the TES bolometer in detail. The signal enters by the tapered Nb microstrip in a waveguide channel. This transmission line crosses a short Si beam, which dominates the thermal conductance between the bolometer and the rest of the detector wafer \citep{rostem-ballistic}. (The long, folded legs that carry the voltage bias do not add significantly to the thermal conduction.) The short Si beam, which is made from the same device layer as the stripline dielectric, is \unit{5}{\mu m} thick. The width of the beam was targeted (measured) to be \unit{10}{\mu m} (9.5--\unit{9.7}{\mu m}), which was chosen to provide \unit{13.2}{pW} of cooling at the target TES critical temperature and the bath temperature.  The short Si beam and bias-lead legs suspend the TES island, which is also \unit{5}{\mu m} thick being built from the device layer. On this island are the Pd thermal mass, the MoAu superconducting bilayer, and the TES termination. The TES was designed to operate at a critical temperature of $\Tc=190\mK$. This increase over the original design's $\Tc= 150\mK$ was to accommodate additional bias power range, matching the more stable \unit{150/220}{GHz} detectors \citep{dahal20HF}. 
The rest of the detector wafer is maintained at a bath temperature $T_{\mathrm{bath}}$ of approximately \unit{50}{mK} by the helium dilution refrigerator-based receiver \citep{iuliano2018spie}.

On the TES island, the Pd thermal mass provides the majority of the heat capacity. In the original design (inset of Figure~\ref{fig:TES_architecture}), four distinct Pd regions define a distributed sensor heat capacity and additionally provide the low-impedance sections of the bias choke. In the revision, the four distinct Pd regions were replaced with a unified, contiguous film that primarily serves to set the heat capacity of the TES. 
To obtain a target heat capacity of \unit{3.7}{pJ/K} at $\Tc=190\mK$, an  area of \unit{45945}{\mu m^2} of Pd was deposited to a thickness of \unit{400}{nm}.\footnote{This assumes a Pd specific heat of $9.4\,\mathrm{mJ/mol/K^2}$ from \cite{kittel96}.}

The MoAu bilayer is the same as in the original design, incorporating the ``zebra-stripe'' normal-metal meander that helps stabilize the TES \citep{Staguhn2004, Ullom-zebra}. Unlike the original, the upgraded design incorporates an additional metallic contact extending the zebra stripes to the Pd. This contact further unifies the thermal circuitry of the island, lumping the electronic heat capacity into a single element. 

\subsection{Revised TES Bolometer Termination}
The bolometer signal termination was revised to incorporate a stepped-impedance Nb-PdAu termination similar to that used for the magic tee and crossover, whereas the original bolometer (inset of Figure~\ref{fig:TES_architecture}) used a distributed quarter-wave short and a sub-wavelength PdAu resistive line for signal termination. The decision to transition to the stepped-impedance design allowed the same basic modeling approach to be used for the TES, magic tee, and crossover terminations, and it addressed concerns about sensitivity of the prior design to misalignment and membrane stresses, described below. 

\subsubsection{Design of Original Termination}
Electromagnetic modeling of this structure needs to be carried out in three dimensions including the surrounding cavity, which has three openings (hereafter called ``ports'') -- one for the incoming microstrip carrying optical power and two for the bias leads traveling to the bolometer island (top-middle, bottom-left, and bottom-right in Figure~\ref{fig:TES_architecture}). 
Modeling of the original TES cavity used the mirror symmetry of the structure to reduce the simulation volume, allowing consideration of only two ports: the signal input and one of the bias leads. This is valid as long as the bolometer island remains planar and there is no misalignment in bonding the backshort with the detector wafer. While the microwave input at the top of Figure \ref{fig:TES_architecture} drives the TES cavity symmetrically, bowing of the TES bolometer structures due to membrane stresses or misalignment of elements in hybridization of the TES enclosure can potentially break this symmetry. Differential modes 
in the structure are not captured in the symmetric simulation. 
Only symmetric modes in which waves are in phase between the bias ports would appear in the model. Microwave power from differential modes is not absorbed on the bolometer island, reducing optical coupling efficiency.

This approach largely followed the successful simulation strategy used in modeling the inputs and readout signals for the 40~GHz sensor, where the wavelength is relatively large compared with the cavity dimensions. Because the size of the enclosing cavity for the TES bolometer is set mostly by the thermal circuit, it remains the same between frequency bands. The smaller wavelength for the 90~GHz and 150/220~GHz detectors compared with the cavity dimensions made this aspect of the design more challenging. In reviewing the sensitivity of the CLASS 150/220 GHz pixel design with the electromagnetic simulations it was realized that the mirrored, 2-port model suppressed coupling to differential modes in the bias circuit. The model was refined to treat the integrated circuit and its packaging structures as a full 3-port structure. This model 
was used to confirm that excitation of the cavity did not lead to trapped mode resonances~\citep{Morgan2013} that could 
cause a loss of efficiency. The 150/220 GHz detector circuitry and its enclosure successfully addressed these considerations through the arrangement of the circuitry within the cavity~\citep{dahal20HF} and changes to the bias line filtering detailed below. 

\subsection{Voltage Bias Line Routing and Filtering}

The routing and filtering of the TES bias leads were revised in the upgraded design. 
In the original bias filter design, the structure resided on the TES island. The new bias filter spans the long, folded legs from the bolometer island to the cavity wall. In addition, the revised design has a larger stop-band~\citep{U-yen-Filters} and works in concert with the TES cavity enclosure as one of several means of mitigating potential “blue-leak” radiation threats.

\begin{figure*}
\centering
\includegraphics[width=\textwidth]{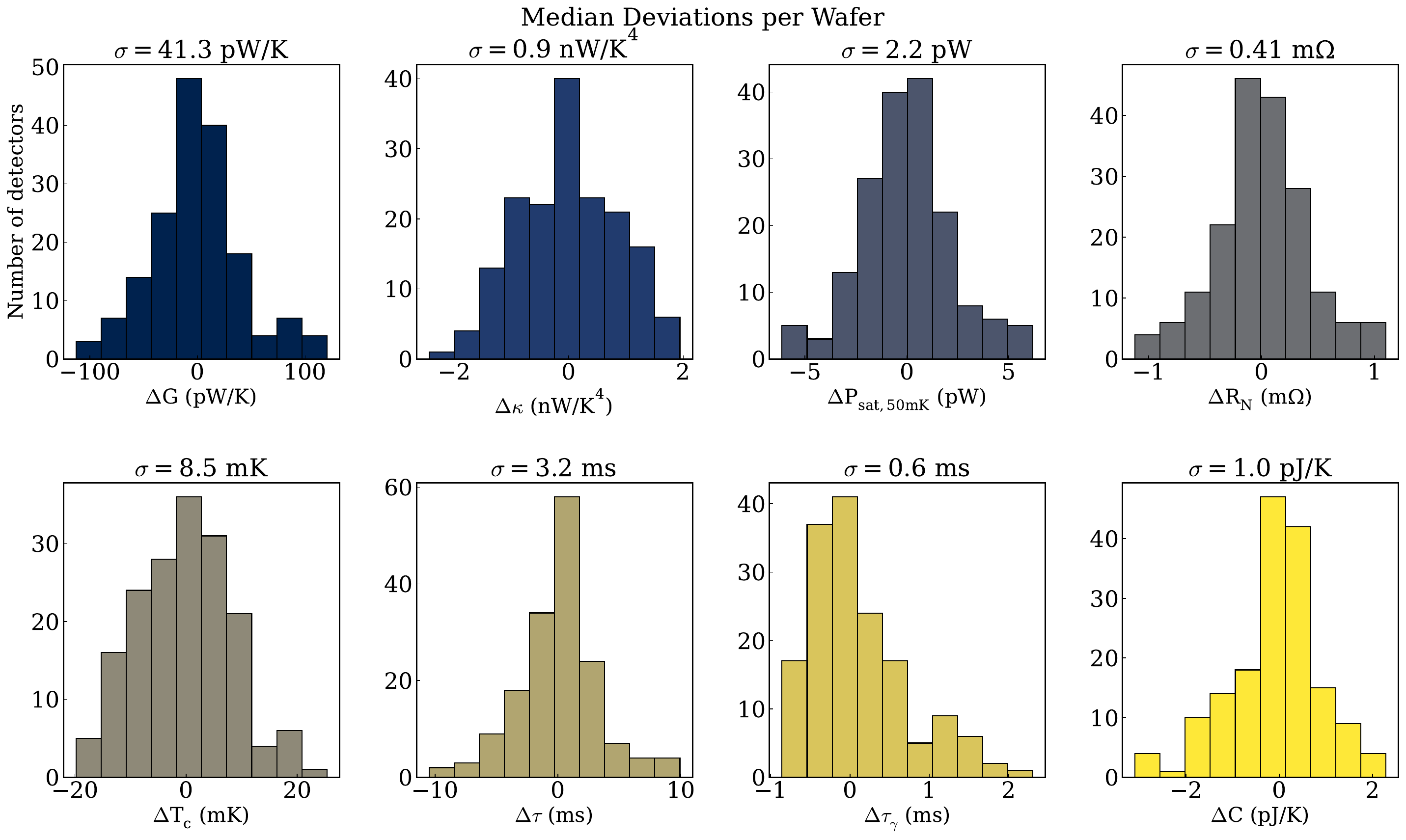}
\caption{Distributions of TES electrothermal parameters relative to their per-wafer median value: thermal conductance ($G$), prefactor ($\kappa$), saturation power for $T_\mathrm{ bath}=50\mK$ ($P_\mathrm{sat}$), normal resistance ($R_N$), critical temperature ($\Tc$), thermal time constant ($\tau$), optical time constant ($\tau_\gamma$), and heat capacity ($C$). Estimation of these parameters is described in Section~\ref{sec:electrothermal}. Outliers beyond $3\sigma$ from the per-wafer median of the underlying distributions have been excluded. Subplot titles show the standard deviation of the histograms. Per-wafer median values for all parameters are given in Table \ref{tab:parameters}. These distributions can be compared to previously published results for the original W-band array (Figure~7 in \cite{dahal18spie}) and the Q-band array (Figure~4 in \cite{appel14spie}). The TES parameters in the upgraded W-band wafers show greater uniformity than those in the 
original wafers but do not match the consistency achieved by selecting individual pixels from Q-band wafers.
}
\label{fig:histograms}
\end{figure*}

\section{Data}
\label{sec:data}

Several datasets were used to characterize the upgraded \unit{90}{GHz} detectors. In-lab non-optical (``dark'') data were used to measure several basic electrothermal parameters (Section \ref{sec:electrothermal}) and phonon noise (Section \ref{sec:noise}). In-lab optical data were used to measure the passbands (Section \ref{sec:bandpass}). Based on cosmological survey data taken from August to October~2022, we estimated optical time constants (Section \ref{sec:electrothermal}) and noise spectra (Section \ref{sec:noise}) for 10-minute data packages. From the survey data, we removed packages for which (1) detectors had anomalous optical response (as indicated by an unusual polarization modulator emission signal), (2) the noise equivalent power (NEP) fluctuated significantly over a contiguous span of data ($<24\,\mathrm{hours}$), (3)  wind speeds exceeded \unit{6}{m\,s^{-1}}, or (4) optical loading was unusually low ($P_\gamma<3.8\,\mathrm{pW}$) or high ($P_\gamma>5.5\,\mathrm{pW}$). We also excluded all detectors that did not have an orthogonally polarized counterpart (i.e., ``unpaired'' detectors).\footnote{Estimating detector NEP using the difference of data from two orthogonally-polarized paired detectors mitigates contributions from noise common to the pair. However, we have found that such  common-mode noise does not impact sensitivity estimates at \unit{90}{GHz} \citep{clearyphdthesis}.} This selection of the survey data is meant to best capture the essential characteristics of the upgraded detectors and some aspects of the receiver (e.g., optical efficiency) rather than characterize the performance of the instrument throughout the survey and under all conditions that will ultimately affect CLASS polarization maps.  Finally, we derived optical efficiencies (Section \ref{sec:efficiency}) from observations of Jupiter carried out in the austral spring of 2022 \citepalias{nunez23asc}.

\section{Electrothermal Parameters}
\label{sec:electrothermal}

In this section we present in-lab and on-sky measurements of the electrothermal properties of the upgraded TES bolometers.  Table~\ref{tab:parameters} shows a summary of the parameters. In addition, Appendix \ref{app:parameter index} contains an index of parameters and relevant equations.
Preliminary values, derived from in-lab measurements, were reported in \citetalias{nunez22spie}. To these results, we append measurements of the optical time constant ($\tau_\gamma$), thermal time constant ($\tau$), and heat capacity ($C$).  Figure ~\ref{fig:histograms} shows the distributions of these parameters for the upgraded wafers relative to their per-wafer median values given in Table \ref{tab:parameters}. In Appendix \ref{app:Detector Property Spatial Variation}, we explore the spatial distribution of these properties across the wafer. As we shall see, the measured values of several device properties diverge from their design targets. In Section~\ref{ssec:Alternative Scenarios} we explore alternative measurement and theory scenarios that could explain these discrepancies. Fortunately, the properties most critical to detector performance, such as saturation power, noise power, optical efficiency, optical time constant, and stability, are scenario independent.

\subsection{Normal Resistance and Saturation Power}
\label{subsec:Rn and Psat}

Several of these parameters were derived using ``$I$--$V$ curves'' showing the TES current ($I_\mathrm{TES}$) as a function of TES voltage ($V_\mathrm{TES}$).
The voltage bias is ramped down in steps, driving the TES from a normal resistor state to a superconducting state.  We convert from the output MCE digital-to-analog feedback units and voltage bias to $I_{\mathrm{TES}}$ and $V_{\mathrm{TES}}$ following \S~4.1 of ~\cite{appel22}.  The normal resistance of the TES ($R_\mathrm{N}$) is given by the inverse slope of the normal branch of the $I$--$V$ curve.  The saturation power ($\Psat$) is the approximate optical loading above which the TES biased on its superconducting transition is driven normal. We estimate the saturation power as the amount of bias power ($P_J=I_{\mathrm{TES}}\times V_{\mathrm{TES}}$) required to maintain the TES at its superconducting critical temperature with a resistance equal to 70\% $R_\mathrm{N}$ with no radiative load (dark configuration). Except when otherwise stated, $\Psat$ will refer to the saturation power for the lowest possible $\Tbath=50$--$60\mK$.\footnote{For the critical temperatures considered here, $\Tc=150$--$190\mK$, the fractional change in $\Psat$ in the range $\Tbath=50$--$60\mK$ is $\sim1\%$, well below uncertainties.} Table \ref{tab:parameters} shows that we measured $\Psat=8$--$12\pW$ for the upgraded detectors. This is $\sim30\%$ below the target of $13.2\pW$. Figure \ref{fig:histograms} shows that within a module, the saturation powers vary with standard deviation $\pm2.2\pW$ about the median. In Appendix \ref{app:Detector Property Spatial Variation} we discuss how this variation takes the form of gradients across the wafers as opposed to spatially random distributions.

\begin{deluxetable*}{ccccccccc}

\tablecaption{
Electrothermal Parameters
\label{tab:parameters}}
\tablehead{
 & \multicolumn{6}{c}{Upgraded Wafers} \\
\cmidrule{2-7} \vspace{-0.25cm}
& & & & & & & Original \\
\vspace{-0.25cm} & \multicolumn{1}{c}{1} & \multicolumn{1}{c}{2} & \multicolumn{1}{c}{3} & \multicolumn{1}{c}{4} & \multicolumn{1}{c}{Total} & \multicolumn{1}{c}{Target} & & \\ & & & & & & & Array\tablenotemark{a}
 }
\startdata
\vspace{-0.27cm} Thermal Conductance  &&&&&&&&\\
\vspace{-0.27cm} & $269_{-44}^{+62}$ & $229_{-27}^{+25}$ & $257_{-45}^{+27}$ & $304_{-53}^{+46}$ & $255_{-42}^{+55}$ & 280 
& 452 \\ 
$G~\mathrm{[pW/K]}$&&&&&&&& \\\hline
\vspace{-0.27cm} Thermal Conductance Prefactor  &&&&&&&& \\
\vspace{-0.27cm}  & $16.1_{-0.72}^{+0.64}$ & $16.9_{-1.0}^{+1.2}$ & $13.9_{-0.94}^{+1.0}$ & $15.2_{-0.98}^{+0.83}$ & $15.2_{-0.98}^{+0.83}$ & 10 
& 24.9\\
$\kappa~\mathrm{[nW/K^4]}$  &&&&&&&& \\\hline
\vspace{-0.27cm} Saturation Power &&&&&&&&\\
\vspace{-0.27cm} & $10.7_{-2.1}^{+3.3}$ & $8.4_{-1.3}^{+1.5}$ & $10.7_{-2.6}^{+1.4}$ & $12.7_{-2.5}^{+2.6}$ & $10.1_{-2.3}^{+3.1}$ & 13.2 
& 18.4\\
$P_{\mathrm{sat}}~\mathrm{[pW]}$  &&&&&&&&\\\hline
\vspace{-0.27cm} Normal Resistance  &&&&&&&& \\
\vspace{-0.27cm} & $12.7_{-0.3}^{+0.5}$ & $11.0_{-0.6}^{+0.5}$ & $10.2_{-0.3}^{+0.2}$ & $10.7_{-0.3}^{+0.4}$ & $10.8_{-0.6}^{+1.8}$ & 12 
& 11.0\\
$R_\mathrm{N}~[\mathrm{m\Omega}]$ &&&&&&&& \\ \hline
\vspace{-0.27cm} Critical Temperature &&&&&&&& \\
\vspace{-0.27cm} & $163_{-7}^{+10}$ & $151_{-10}^{+8}$ & $165_{-12}^{+8}$ & $171_{-7}^{+6}$ & $161_{-10}^{+12}$ & 190 
& 167\\
$T_\mathrm{c}~\mathrm{[mK]}$ &&&&&&&& \\ \hline
\vspace{-0.27cm} Thermal Time Constant &&&&&&&& \\
\vspace{-0.27cm} & $17.3_{-2.9}^{+4.3}$ & $15.8_{-3.9}^{+2.2}$ & $17.0_{-2.5}^{+2.0}$ & $12.6_{-3.8}^{+3.0}$ & $16.0_{-4.6}^{+2.8}$ & 13.2 
& 7 \\
$\tau~\mathrm{[ms]}$ &&&&&&&& \\ \hline
\vspace{-0.27cm} Optical Time Constant &&&&&&&& \\
\vspace{-0.27cm} & $3.5_{-0.3}^{+0.8}$ & $4.7_{-0.6}^{+1.4}$ & $4.6_{-0.4}^{+1.9}$ & $3.2_{-0.3}^{+1.3}$ & $4.2_{-0.9}^{+1.0}$ & -- 
& 2.1\\
$\tau_{\gamma}~\mathrm{[ms]}$ &&&&&&&& \\ \hline
\vspace{-0.27cm} Heat Capacity &&&&&&&& \\ 
\vspace{-0.27cm} & $4.8_{-1.1}^{+1.2}$ & $3.8_{-1.2}^{+0.5}$ & $4.5_{-1.3}^{+0.6}$ & $3.8_{-1.0}^{+1.2}$ & $4.1_{-1.2}^{+0.9}$ & 3.7 
& 4\\
$C~\mathrm{[pJ/K]}$ &&&&&&&& 
\enddata

\tablecomments{Median values and 68\% intervals of key electrothermal parameters for the four upgraded wafers (numbered as shown in~\ref{subfig:W1_in_field}) and the originally deployed W1 array.  The ``Total'' column shows the median values across the four upgraded wafers, and the ``Original Array'' column shows the median values across the full seven-wafer array that was originally deployed.}
\tablenotetext{a}{Values reported in \citetalias{dahal22}.  $P_\mathrm{sat}$ values published therein were calculated at 80\% $R_\mathrm{N}$, and 70\% $R_\mathrm{N}$ in this work. For the upgraded detectors the measured $P_\mathrm{sat}$ drops by $\sim$2\%  when calculated at 70\% $R_\mathrm{N}$ compared to 80\% $R_\mathrm{N}$. }
\end{deluxetable*}

\subsection{Critical Temperature and Thermal Conductivity}
\label{subsec:Tc and G}

Using the same $I$--$V$ curve approach described above, we measured saturation powers $P_{\mathrm{sat},i}$ corresponding to a wide range of bath temperatures $60\mK<T_{\mathrm{bath},i}<250\mK$ \citepalias{nunez22spie}. With these measurements, we  simultaneously determined $T_\mathrm{c}$ and the thermal-conductance prefactor $\kappa$ using the relation~\citep{TES-chapter}:
\begin{equation}
    P_{\mathrm{sat},i} = \kappa \left(T_\mathrm{c}^n - T_{\mathrm{bath},i}^n\right) \mathrm{.}
    \label{eqn:Psat}
\end{equation}
In keeping with the assumption that the short Si beam connecting the TES to the bath operates in the ballistic phonon limit, we assumed $n=4$. From Table \ref{tab:parameters} we see that the median critical temperatures for the upgraded modules were measured to be $\Tc=150$--$170\mK$, about $15\%$ lower than the target of $\Tc=190\mK$. Because $\Psat\propto\Tc^n$, the lower $\Tc$ implies a $\sim50\%$-lower saturation power, all other parameters being on target. Given that saturation power is only $\sim30\%$ below the target implies that the prefactor $\kappa$ is higher than designed (though see Section \ref{ssec:Alternative Scenarios}). The prefactor $\kappa$ is a function of the geometry of the short Si beam \cite[Figure \ref{fig:TES_architecture},][]{rostem-ballistic}. Table \ref{tab:parameters} shows that the measured $\Psat$ and $\Tc$ imply prefactors in the range $\kappa = 14$--$17\,\mathrm{nW\,K^{-4}}$ compared to the design value of $10\,\mathrm{nW\,K^{-4}}$. Figure \ref{fig:histograms} shows that within a wafer, the standard deviation of the $\kappa$ values is small, approximately $0.9\,\mathrm{nW\,K^{-4}}$, so the discrepancy between the measured value and the design value cannot be attributed to a statistical fluctuation in the measurement. Finally, the thermal conductance ($G$) of the TES is given by~\citep{TES-chapter}:
\begin{equation}
    G = \frac{dP}{dT}\Bigr|_{\substack{T_\mathrm{c}}} = n \kappa T_\mathrm{c}^{n-1} \mathrm{.}
    \label{eqn:G}
\end{equation}
Because $\Tc^n\gg \Tbath^n$, the thermal conductance is well approximated as $G\approx n\Psat/\Tc$. Therefore, we would expect the measured $G$ to be approximately $10\%$ low compared to the design target. This is indeed the case with measured values $G=230$--$300\,\mathrm{pW/K}$ versus the designed $280\,\mathrm{pW/K}$. Figure \ref{fig:histograms} shows that the range of values per wafer has a standard deviation of $41.3\,\mathrm{pW/K}$.

\subsection{Time Constants and Heat Capacity}
\label{subsec:tau and C}

\begin{figure}
\centering
\includegraphics[width=\linewidth]{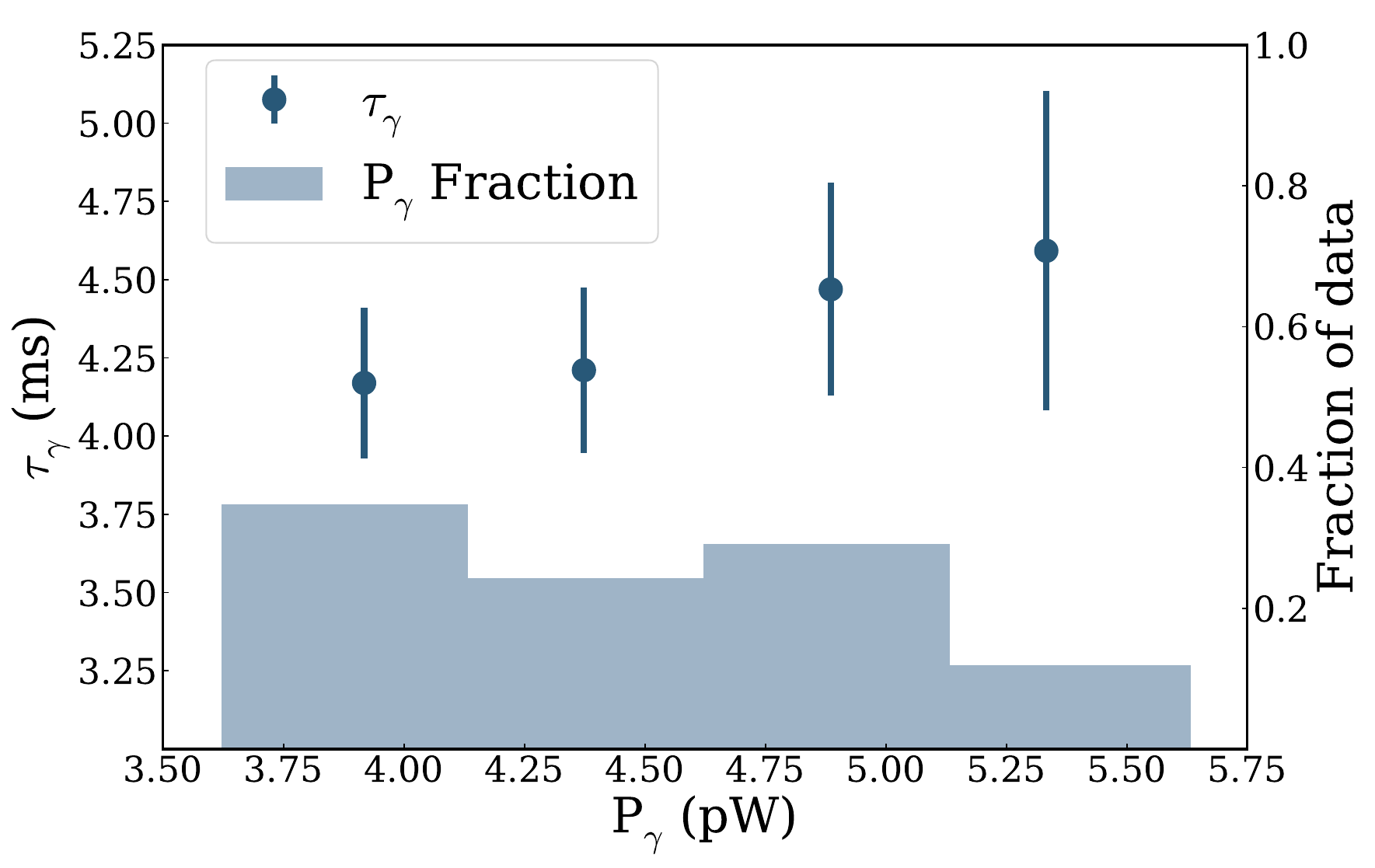}
\caption{The optical time constants of the upgraded detectors as a function of optical loading $P_\gamma$. Increasing optical loading results in a decrease in bias power and therefore electrothermal feedback. The drop in feedback is consistent with the marginal trend of slowing of the detectors with increasing $P_\gamma$.}
\label{fig:tau_vs_Pload}
\end{figure}

To measure the on-sky optical time constant ($\tau_\gamma$), we fit for the single-pole filter that minimizes the hysteresis of the periodic modulator emission signal in the TODs, as in \cite{appel19}. Table \ref{tab:parameters} shows that the median $\tau_\gamma$ per wafer lies in the range $3$--$4\,\mathrm{ms}$. This time constant is shorter than the thermal time constant ($\tau=C/G$) due to electrothermal feedback. The speed up increases with both bias power $P_J$ and the logarithmic derivative of resistance with respect to temperature ($\alpha$) \citep{TES-chapter}. Figure~\ref{fig:tau_vs_Pload} shows $\tau_\gamma$ as a function of optical loading ($P_\gamma$). Because $\Psat = P_\gamma + P_J$, an increase in optical loading results in a decrease in bias power. This decreases the electrothermal feedback, which results in a longer optical time constant, consistent with the trend of increasing $\tau_\gamma$ with increasing $P_\gamma$ in Figure \ref{fig:tau_vs_Pload}.

\cite{appel22} describes how to use detector responsivity measurements (derived from $I$--$V$ curves) to obtain the thermal time constant $\tau$ from the optical time constant $\tau_\gamma$. (See Equation 26 of \citealt{appel22} and associated discussion for details.) Applying this formalism, Table \ref{tab:parameters} shows that we found median thermal time constants in the range $\tau=13$--$17\,\mathrm{ms}$. Combining this result with measured thermal conductivities gives median heat capacities per wafer in the range $C=3.8$--$4.8\,\mathrm{pJ/K}$ (Table \ref{tab:parameters}) with an RMS variation per wafer of $\sigma_{\Delta C}=1\,\mathrm{pJ/K}$ (Figure \ref{fig:histograms}). As discussed in Section \ref{sec:design}, the TES heat capacity is dominated by the $400\,\mathrm{nm}$-thick Pd covering most of the TES island. The heat capacity of the design listed in Table~\ref{tab:parameters} ($3.7\,\mathrm{pJ/K}$) was computed for the target $\Tc=190\mK$. Because the heat capacity of normal metals is proportional to temperature, the expected heat capacity for the measured $\Tc=160\mK$ is revised downwards to $C\approx3.1\,\mathrm{pJ/K}$. Therefore, the measured heat capacity estimate is higher on average by approximately $1\,\mathrm{pJ/K}$ than the expectation. In the next section, we investigate whether systematic errors in the measurements or deviations in the assumptions presented above can reconcile discrepancies between our baseline parameter estimates and the design targets.

\subsection{Alternative scenarios}
\label{ssec:Alternative Scenarios}

In the previous section, we found a discrepancy between the measured and targeted thermal conductance prefactor $\kappa$, which is in principle a function of the dimensions of the short mono-crystaline Si beam connecting the TES island to the rest of the detector wafer. We also measured an elevated heat capacity relative to that expected from the $400\,\mathrm{nm}$-thick Pd layer covering most of the island. In this section, we explore two alternative scenarios and whether they could reconcile the measured values with the design targets.

First, we consider what changes if we relax the assumption that ballistic photon transport dominates the thermal conductance across the short Si beam. Non-idealities would result in $n<4$ in Equations \ref{eqn:Psat} and \ref{eqn:G}.  Because $\Psat$ is measured by the bias power under dark conditions, it is independent of the model assumptions. Given that $\Psat\approx \kappa \Tc^n$ and $\Tc < 1$, decreasing $n$ would require decreasing $\kappa$. Setting $n = 3.77$ would shift the average measured $\kappa$ from $15.2\,\mathrm{nW/K^4}$ to the target $10\,\mathrm{nW/K}^n$.\footnote{Unfortunately, the $n$ and $\kappa$ parameters are significantly covariant in a simultaneous fit to $IV$ data, so we are restricted to fixing $n$ and fitting for $\kappa$ or vice-versa.} Therefore relaxing the assumption on $n$ can resolve the $\kappa$ discrepancy. Unfortunately, it cannot simultaneously explain why the estimated heat capacity is 25\% above expectation. As described in Section~\ref{subsec:tau and C}, we estimated the heat capacity from measurements as $C=\tau G$. The thermal conductivity $G \approx n\Psat/\Tc$, meaning the 5\%  decrease in $n$ considered here would marginally reduce the $C$ discrepancy. The thermal time constant $\tau$ is to first order inversely dependent on $n$ through $G$ \citep{appel22}, canceling some of these gains.

Second, we consider whether a systematic error in the measurement of $\Tc$ could resolve the discrepancies. Variations between thermometers on different detector modules during $I$--$V$ curve measurements would allow for approximately $10\mK$ systematic error. As in the previous paragraph, we take $\Psat\approx\kappa\Tc^n$ to be a fixed, measured quantity. Therefore, if the measured $\Tc$ were below the actual $\Tc$, then the measured $\kappa$ would be higher than the actual $\kappa$, a change that would bring the actual $\kappa$ into better agreement with the design value. If the critical temperature were $\Tc\approx170\mK$ instead of the measured $\sim160\mK$, then the inferred $\kappa$ would shift to $\kappa\approx12\,\mathrm{nW/K^4}$, reducing the discrepancy from the target value from $50\%$ of the expected value to $20\%$.  Next, we consider what a $10\mK$ increase in $\Tc$ would mean for the discrepancy between measured ($4.1\,\mathrm{pJ/K}$) and expected  ($3.1\,\mathrm{pJ/K}$) heat capacity. Because the heat capacity of metals increases linearly with temperature, the expected heat capacity would increase to approximately $3.3\,\mathrm{pW/K}$. If we treat the thermal time constant $\tau$ as a measured, fixed quantity, then the heat capacity inferred from measurements $C=\tau G= \tau (n\Psat/\Tc)$ would decrease to approximately $3.9\mathrm{\pW/K}$. Therefore, the discrepancy between measured and expected heat capacity decreases from 30\% of the expected value to 15\%. In summary, a $+10\mK$ shift in critical temperature significantly reduces the discrepancies between the design targets and measured $\Tc$, $\kappa$, and $C$. As an aside, measurements of noise (Section \ref{sec:noise}) do not strongly constrain $\Tc$ given a fixed $\Psat$: our noise measurements can accommodate either $\Tc=160\mK$ or $\Tc=170\mK$. One final observation that is consistent with $\Tc=170\mK$: after depositing the TES MoAu bilayer, ``witness'' samples deposited along side the detectors yielded $\Tc=165$--$182\mK$. These were measured in a four-wire configuration at NASA, independently from other measurements described here. However, the TES bilayers undergo further processing, including the addition of the gold meander and niobium leads, that could change the average $\Tc$ of the finished TESs from that of the witness samples. In particular, film stress is likely to be a major factor in $\Tc$ change with the $\Tc$ of a TES changing between a solid frame and a released membrane.  

In summary, having a lower $n$ or a higher $\Tc$ are alternative scenarios that bring the $\kappa$ inferred from measurements into better agreement with the design. And the $\Tc$ adjustment is more effective at solving the discrepancy for heat capacity. Of course, the reality is likely a more general scenario that combines deviations in $\kappa$, $\Tc$, and $n$ from the baseline assumptions of Section~\ref{sec:electrothermal}. Regardless, the key observational parameters---$\Psat$, NEP/NET, $\eta$, $\tau$, etc.---are measured values and independent of the model assumptions.

\section{Bandpass Measurements}
\label{sec:bandpass}
 
Bandpass measurements were made using a Fourier Transform Spectrometer (FTS) utilizing a polarizing Martin-Puplett interferometer \citep{FTS}. The FTS has a movable mirror that sweeps through a 150~mm range at 0.5~mm/s, resulting in a resolution of $\sim$1.6~GHz. The input signal of the FTS is a wide-band thermal source with a temperature of $\sim1000\degree\mathrm{C}$.
The output signal of the FTS, measured by the detectors, is modulated with a chopper at 21~Hz. The bandpass is given by the real component of the Fourier transform of the interferogram.

\begin{figure}
\centering
\includegraphics[width=.99\linewidth]{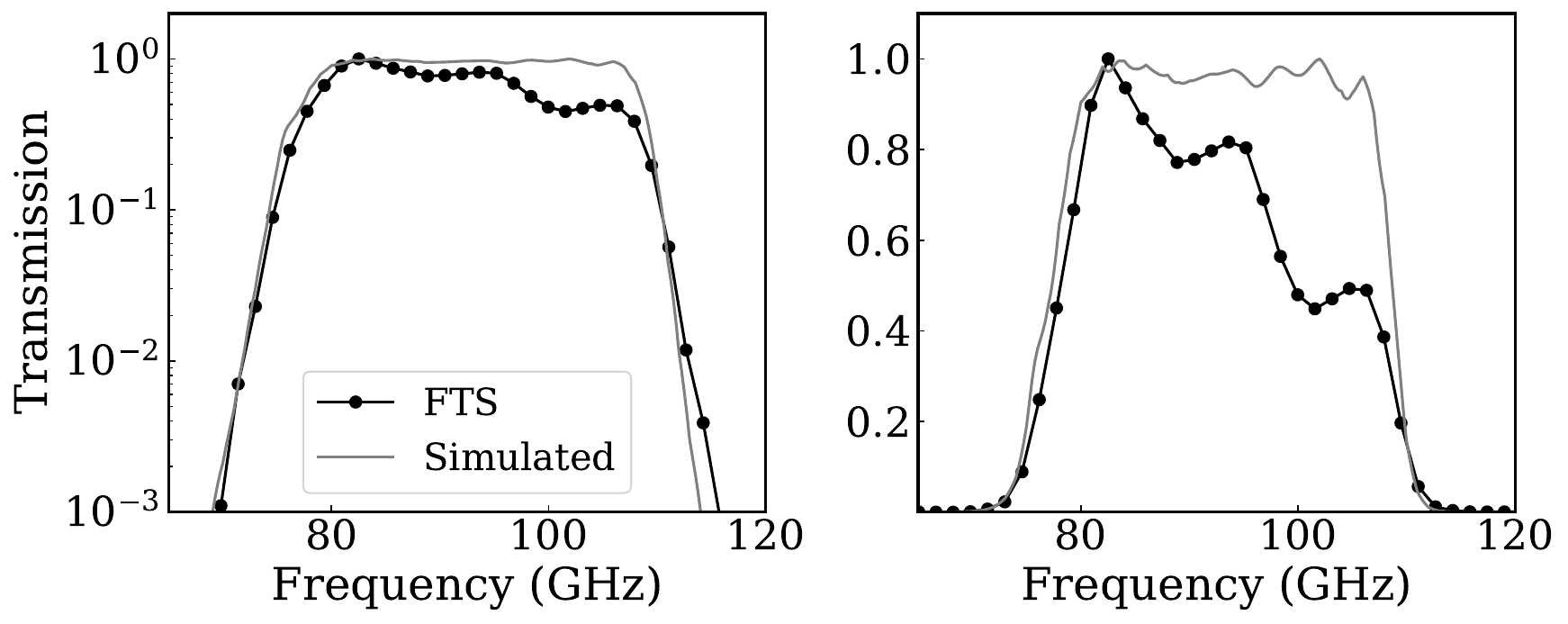}
\caption{Bandpass FTS measurement and simulation. On the left in logarithmic scale and on the right in linear scale. The measured bandpass edges agree well with the simulation. In the text, we present arguments that the decrement in transmission on the high-frequency half of the bandpass is a measurement artifact. 
}
\label{fig:bandpass}
\end{figure}
The measured response plotted in Figure~\ref{fig:bandpass} is the inverse-variance-weighted average over 93 bolometers that yielded high-quality interferograms and bandpasses across wafers 1, 2, and 3.\footnote{Wafer 4 was not delivered in time for the FTS measurement.} We also simulated the bandpass for the upgraded detector architecture using Ansoft HFSS.\footnote{ANSYS 2019 R1, https://www.ansys.com/products/electronics/ansys-hfss} The measured bandpass edges are in good visual agreement with the simulation; the data, however, show an approximately $50\%$ decrement of power in the upper section of the band, especially above $100\GHz$. Measurements made with a second FTS did not show this decrement \citep{Pan2019}. (An observed $\pm1\GHz$ systematic error in our frequency calibration of the second FTS made it less reliable for determining the absolute values for bandpass edges.) Furthermore, the power measured from Jupiter (Section \ref{sec:efficiency}) is too high if the smaller bandwidth implied by the measurement is assumed. Therefore, we conclude that discrepancies observed in-band between measured and simulated responses are likely due to optical effects related to the test setup. 

In Table \ref{tab:bandpass} we report bandwidths and effective bandcenters for diffuse sources for the simulated bandpass and the measurement shown in Figure~\ref{fig:bandpass}. These are computed following the definitions in Section 3.1 of \citetalias{dahal22}. Uncertainties for the bandwidths (effective center frequencies) are given by the summation in quadrature of the standard errors on the mean and the FTS measurement resolution (half FTS measurement resolution). Due to the likely-artificial power suppression in the measurement, the simulated bandwidths and bandcenters are preferred over the measured quantities.

Before integration into the CLASS detector modules, each 37-pixel detector wafer is hybridized into a stack of silicon wafers. The overall design safeguards against the detection of out-of-band radiation (a.k.a., ``blue leak'', for the common case that the spurious signal is at higher frequency). These safeguards include a photonic choke etched into the wafer that contacts the CLASS module's interface plate to the feedhorns, an indium-sealed degenerately-doped silicon back-short cap that shields the detector circuitry after the antenna probes, and RF filtering on the bias leads of the TES \citep{Denis-fabrication, rostem16spie, dahal18spie}. As was found with previous generations of CLASS detectors, the upgraded detectors show good suppression of power outside the passband: FTS measurements in the range $40$--$300\GHz$ place an upper limit on out-of-band power at $0.7\%$ of the in-band response.

\begin{deluxetable}{lcc}
\tablecaption{ Bandpass Parameters \label{tab:bandpass} }
\tablehead{
\colhead{Parameter}  & \colhead{Simulated} & \colhead{Measured}\\
 & \colhead{(GHz)} & \colhead{(GHz)}}
\startdata
Bandwidths & & \\
Full-Width Half-Power & $31.4\pm1.8$ & $21.5\pm1.8$\\
Dicke & $33.9\pm1.7$ & $32.0\pm1.7$ \\
\hline
Effective Bandcenters & & \\
CMB  & $92.4\pm0.9$ & $90.8\pm0.9$ \\
Dust  & $94.2\pm0.9$ & $92.7\pm0.9$\\
Rayleigh--Jeans  & $92.8\pm0.9$ & $91.2\pm0.9$\\
Synchrotron  & $89.8\pm0.9$ & $88.9\pm0.9$
\enddata
\tablecomments{ For reasons given in the text, we consider bandcenters and bandwidths computed from the simulated bandpass  as more reliable. Measured values were previously reported in \citetalias{nunez23asc}. }
\end{deluxetable}

\section{Optical Efficiency Improvements}
\label{sec:efficiency}

We report telescope optical efficiency improvements based on observations of Jupiter. A full description of the Jupiter observations is given in \citetalias{nunez23asc}. In summary, optical efficiencies were measured by comparing the observed and expected amplitudes of Jupiter in the Rayleigh--Jeans limit, given by:
\begin{equation}
    A_\mathrm{J} = k_\mathrm{B} T_\mathrm{J} \Delta \nu B_{\mathrm{dil}}\,,
\end{equation}

\noindent where $k_\mathrm{B}$ is the Boltzmann constant, $T_\mathrm{J}$ is the Jupiter brightness temperature at 90~GHz (172.8~K) as reported by the WMAP team \citep{bennett13}, $\Delta \nu$ is the observing bandwidth of 31~GHz (consistent with the FWHP of the simulated bandpass in Table \ref{tab:bandpass}), and $B_{\mathrm{dil}}$ is the beam dilution factor given by the ratio of the oblateness-corrected solid angle of Jupiter ($\Omega_\mathrm{J}$) over the convolution of the detector beam with Jupiter ($\Omega_\mathrm{B}$).

\begin{figure}
\centering
\includegraphics[height=1.5in]{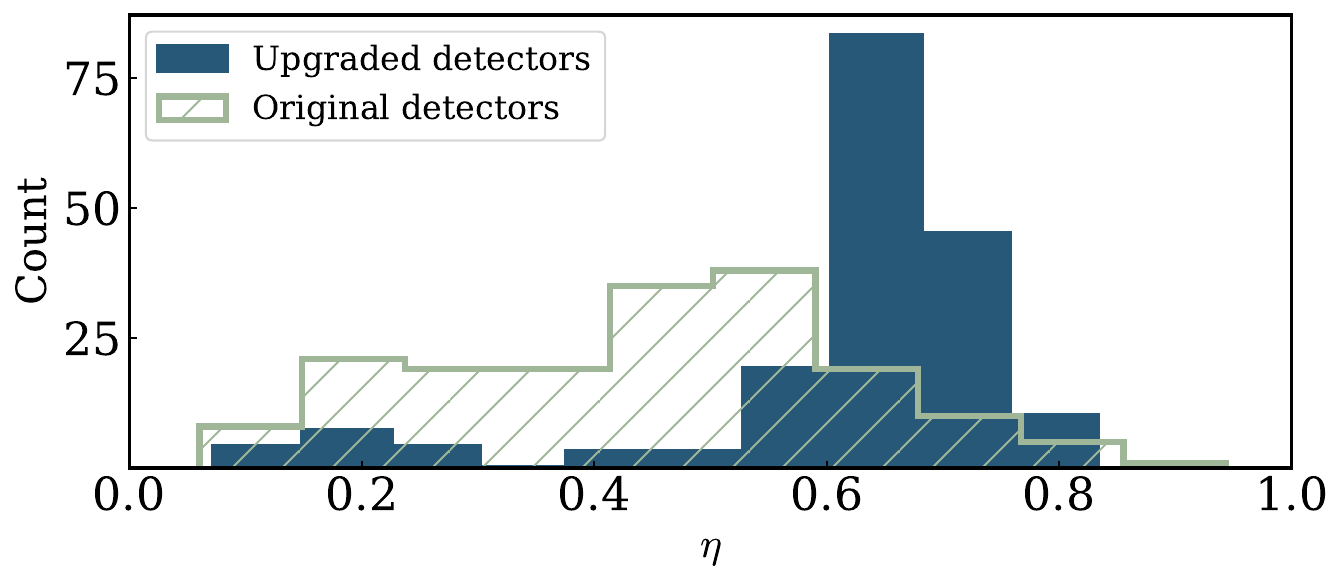}
\caption{Telescope optical efficiencies for the 353 detectors in the upgraded array with optical efficiencies above 5\%. The upgraded detectors provide higher optical efficiency compared to their original counterparts in the upgraded array. (The originals, not included here, that were replaced by the upgraded detectors had poorer performance than those that were retained.) 
Of the sixteen upgraded detectors with optical efficiencies below 0.4, eleven come from wafer 1, which had fabrication issues that were improved for wafers 2--4.}
\label{fig:histogram_optical_effs}
\end{figure}

\begin{deluxetable*}{ccccccccc}

\tablecaption{Optical Efficiency, Yield, and Noise Performance\label{tab:optical-performance}}

\tablehead{
 & \multicolumn{6}{c}{Upgraded Wafers} \\
\cmidrule{2-7} \vspace{-0.25cm}
& & & & & & & Upgraded & Original \\
\vspace{-0.25cm} & \multicolumn{1}{c}{1} & \multicolumn{1}{c}{2} & \multicolumn{1}{c}{3} & \multicolumn{1}{c}{4} & \multicolumn{1}{c}{Total} & \multicolumn{1}{c}{Target} & & \\ & & & & & & & Array &  Array\tablenotemark{a}
 }
\startdata
\vspace{-0.27cm} Telescope Optical Efficiency&&&&&&&& \\
\vspace{-0.27cm} & $0.61_{-0.40}^{+0.07}$ & $0.67_{-0.04}^{+0.03}$ & $0.66_{-0.05}^{+0.06}$ & $0.63_{-0.10}^{+0.08}$ & $0.65_{-0.06}^{+0.06}$ & 0.53 & $0.60_{-0.32}^{+0.10}$ & $0.42_{-0.22}^{+0.15}$ \\
$\eta$ &&&&&&&& \\ 
\vspace{.2cm} Detector Yield\tablenotemark{b}  & 42 (57\%) & 51 (69\%) & 48 (65\%) & 37 (50\%) & 178 (60\%)  & 80\% & 344 (66\%) & 319 (62\%) \\
\vspace{.2cm}Detector $\mathrm{NEP}_\mathrm{dark}$~[$\mathrm{aW}\sqrt{\mathrm{s}}$] & $12.2_{-1.1}^{+2.8}$ & $11.7_{-1.6}^{+2.2}$ & $11.9_{-1.4}^{+2.6}$ &  $13.2_{-1.6}^{+4.2}$ & $12.3_{-1.6}^{+3.3}$ & --  & -- & 24 \\
\vspace{.2cm} Detector Total NEP~[$\mathrm{aW}\sqrt{\mathrm{s}}$] & $32.9_{-8.8}^{+7.8}$ & $32.4_{-4.1}^{+9.8}$ & $35.1_{-2.5}^{+5.8}$ & $35.9_{-6.6}^{+4.9}$ & $34.2_{-5.8}^{+7.6}$  & -- & $34.6_{-6.0}^{+9.9}$ & 47\\
$\mathrm{NET}$~[$\mu\mathrm{K}\sqrt{\mathrm{s}}$] & $165_{-13}^{+314}$ & $149_{-8}^{+9}$& $160_{-8}^{+12}$ & $166_{-18}^{+21}$ & $158_{-12}^{+21}$ & --  & $9.7$ & 19 \\
\enddata
\tablecomments{NEP, NET, and optical efficiency values are given as the median for single detectors with associated 68\% intervals. The exceptions are the NET values in the ``Array'' columns, which are the sensitivities associated with the inverse variance weighted average across all detectors in the arrays.}
\tablenotetext{a}{Values for the original array are from Table~3 of \citetalias{dahal22}.}
\tablenotetext{b}{There are a maximum of 74 detectors per wafer. The ``Total'' and ``Array'' columns have maximums of 298 and 518, respectively.}
\end{deluxetable*}

The telescope optical efficiency\footnote{Telescope optical efficiency is the product of detector efficiency with the absorption and spillover efficiency of the telescope. Unless otherwise specified, abbreviations ``optical efficiency'' and ``efficiency'' refer to this quantity.} denoted by $\eta$ is given by the ratio of the measured amplitude $A_{\mathrm{obs}}$ with the expected amplitude $A_\mathrm{J}$:
\begin{equation}
    \eta = \frac{A_{\mathrm{obs}}}{A_\mathrm{J}} = \frac{A_{\mathrm{obs}} \Omega_{\mathrm{B}}}{ k_\mathrm{B} \Delta\nu T_{\mathrm{J}} \Omega_{\mathrm{J}}}\,.
    \label{eqn:efficiency}
\end{equation} The distribution of telescope optical efficiencies of the currently fielded detectors is shown in Figure~\ref{fig:histogram_optical_effs}. 
Table \ref{tab:optical-performance} gives the median optical efficiency and 68\% confidence region for each upgraded wafer, for the four upgraded wafers together, and for the full detector array with seven wafers.  The optical efficiency across all upgraded detectors is $0.65\pm0.06$.  \cite{iulianophdthesis} estimated the efficiency of the $90\GHz$ receiver optics accounting for only spill to be $0.78$, and $0.58$ when also including absorption and reflection.  The absorption in cryogenic filters and lenses was likely overestimated. A plausible scenario consistent with the upgraded detector telescope optical efficiency of $0.65$ includes spill efficiency of $0.78$, detector efficiency of $0.91$, and telescope absorption and reflection efficiency of $0.91$.
By contrast, Figure~\ref{fig:histogram_optical_effs} shows that the original detectors retained in the upgraded array have a broader distribution of optical efficiencies with a lower median. Altogether, the efficiencies for original and upgraded detectors in the upgraded array are distributed as $0.60^{+0.10}_{-0.32}$. This is an improvement compared with the original array, which had optical efficiencies distributed as $0.42^{+0.15}_{-0.22}$ \citepalias[Table \ref{tab:optical-performance},][]{dahal22}.

Not included in Figure~\ref{fig:histogram_optical_effs} are detectors with optical efficiencies below 5\%. Conversely, the fraction of detectors with $\eta>0.05$ defines the detector yield, which is given in Table \ref{tab:optical-performance}. The table shows that, although the upgraded design improved the efficiency of yielded detectors, the yielded fraction of the upgraded wafers (66\%) did not significantly improve over that of the original array (62\%). Based on electrical and visual inspections of the detectors, we believe the low detector yield for the upgraded wafers is likely due to open and short circuits along the bias leads extending from the detector pixels to the bondpads at the edge of the wafer. For a detailed look at the spatial distribution of detector yield, see Figure~\ref{fig:wafer_plots} of Appendix \ref{app:Detector Property Spatial Variation}.

\begin{figure*}
\centering
\includegraphics[width=.8\linewidth]{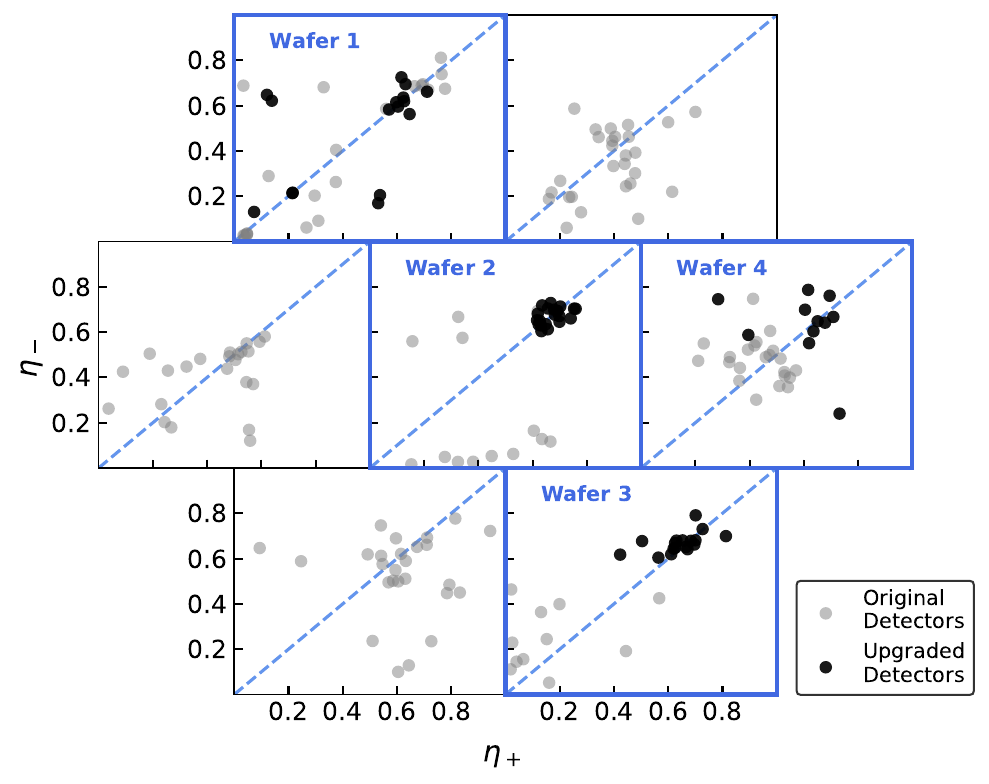}
\caption{Optical efficiency comparison between paired detectors within a pixel for the original and upgraded focal planes. The optical efficiencies for the $+45\degree$ and $-45\degree$ detectors, as labeled in Fig.~\ref{subfig:revised_pixel}, are denoted $\eta_+$ and $\eta_-$. Each subplot represents a wafer on the focal plane.  The subplots with blue edges correspond to the four upgraded wafers of the fielded focal plane shown in Figure~\ref{subfig:W1_in_field}. The originally fielded detectors are shown in gray, and the upgraded detectors are shown in black. In addition to higher efficiencies, the upgraded detectors have less scatter of relative efficiencies within a detector pair.  Fabrication issues with Wafer 1 resulted in a higher number of lower-efficiency detectors; however, these fabrication issues were improved for Wafers~2--4.}
\label{fig:efficiencies}
\end{figure*}

Finally, in Figure~\ref{fig:efficiencies}, we present a comparison of optical efficiencies between the paired $\pm45^\circ$ polarized detectors within a pixel. This comparison for the original detectors was previously published in Figure 7 of \citet{datta23}. Because paired detectors are colocated within a square centimeter on a wafer, it is unlikely that discrepancies in optical coupling between paired detectors arise due to variations in fabrication that are more pronounced between wafers or across the full area of a wafer. More likely, intra-pair discrepancies arise due to a design's susceptibility to the more modest variation in depositions, lithography, and  stress/strain on the TES membrane. 
We note that the lowest efficiency detectors (also visible in Figure \ref{fig:histogram_optical_effs}) reside on wafer 1, which suffered from fabrication issues that were improved for wafers 2--4.\footnote{Fabrication issues included four broken membranes, wiring shorts, and a problem with the liftoff process for the magic tee and dummy cross-over terminations.}

\section{Noise Performance}
\label{sec:noise}

We quantify the noise performance by measuring the noise-equivalent power (NEP).  The total detector NEP is the quadrature sum of contributions from the incoming light ($\mathrm{NEP}_\gamma$) and from all other sources ($\mathrm{NEP}_\mathrm{dark}$):
\begin{equation}
 \left(\mathrm{NEP}\right)^2 = \left(\mathrm{NEP}_\gamma\right)^2 + \left(\mathrm{NEP}_\mathrm{dark}\right)^2 \,. 
 \label{eqn:total_NEP}
\end{equation}
$\mathrm{NEP}_\mathrm{dark}$ is determined in lab by measuring noise spectra with the cryostat in a dark configuration.  In-lab measurements yielded a median $\mathrm{NEP}_\mathrm{dark}$ of 
$12.3\pm3.3\,\mathrm{aW}\sqrt{\mathrm{s}}$ \citepalias{nunez22spie}.  $\mathrm{NEP}_\mathrm{dark}$ consists of contributions due to phonon noise, SQUID noise, and Johnson noise.  For the CLASS \unit{90}{GHz} detectors, $\mathrm{NEP}_\mathrm{SQUID} \approx 3 \, \mathrm{aW}\sqrt{\mathrm{s}}$,  and $\mathrm{NEP}_\mathrm{Johnson}$ is suppressed due to electrothermal feedback.  $\mathrm{NEP}_\mathrm{phonon}$ is given by:
\begin{equation}
    \mathrm{NEP}_\mathrm{phonon} = \sqrt{2 k_B T_\mathrm{c}^2 G F_\mathrm{link}}\,.
\end{equation}
In the case of ballistic phonon transport, $F_\mathrm{link}$ is given by $(1+ (T_\mathrm{bath}/T_\mathrm{c})^{n+1})/2$ \citep{Boyle}. For CLASS, $T_\mathrm{bath}\approx 50\,\mathrm{mK}$ and $F_\mathrm{link} \approx 1/2$.  Using the median $T_\mathrm{c} = 160 \, \mathrm{mK}$ and $G = 255 \, \mathrm{pW}\, \mathrm{K}^{-1}$ for the four upgraded wafers, we find $\mathrm{NEP}_\mathrm{phonon} \approx 9.5 \, \mathrm{aW}\sqrt{\mathrm{s}}$. In the allowed alternative scenario of $T_\mathrm{c}=170\mK$ explored in Section \ref{ssec:Alternative Scenarios}, the predicted $\mathrm{NEP}_\mathrm{phonon}$ is increased by only a few percent because $\mathrm{NEP}_\mathrm{phonon}\propto\sqrt{\Psat T_\mathrm{c}}$, with $\Psat$ being a fixed, measured quantity. Therefore the fact that the median measured $\mathrm{NEP}_\mathrm{dark}$ is higher than predicted by the electrothermal parameters favors neither the nominal nor alternative scenarios, and there is likely some undetermined additional contribution to $\mathrm{NEP}_\mathrm{dark}$.

\begin{figure}
\centering
\includegraphics[width=\linewidth]{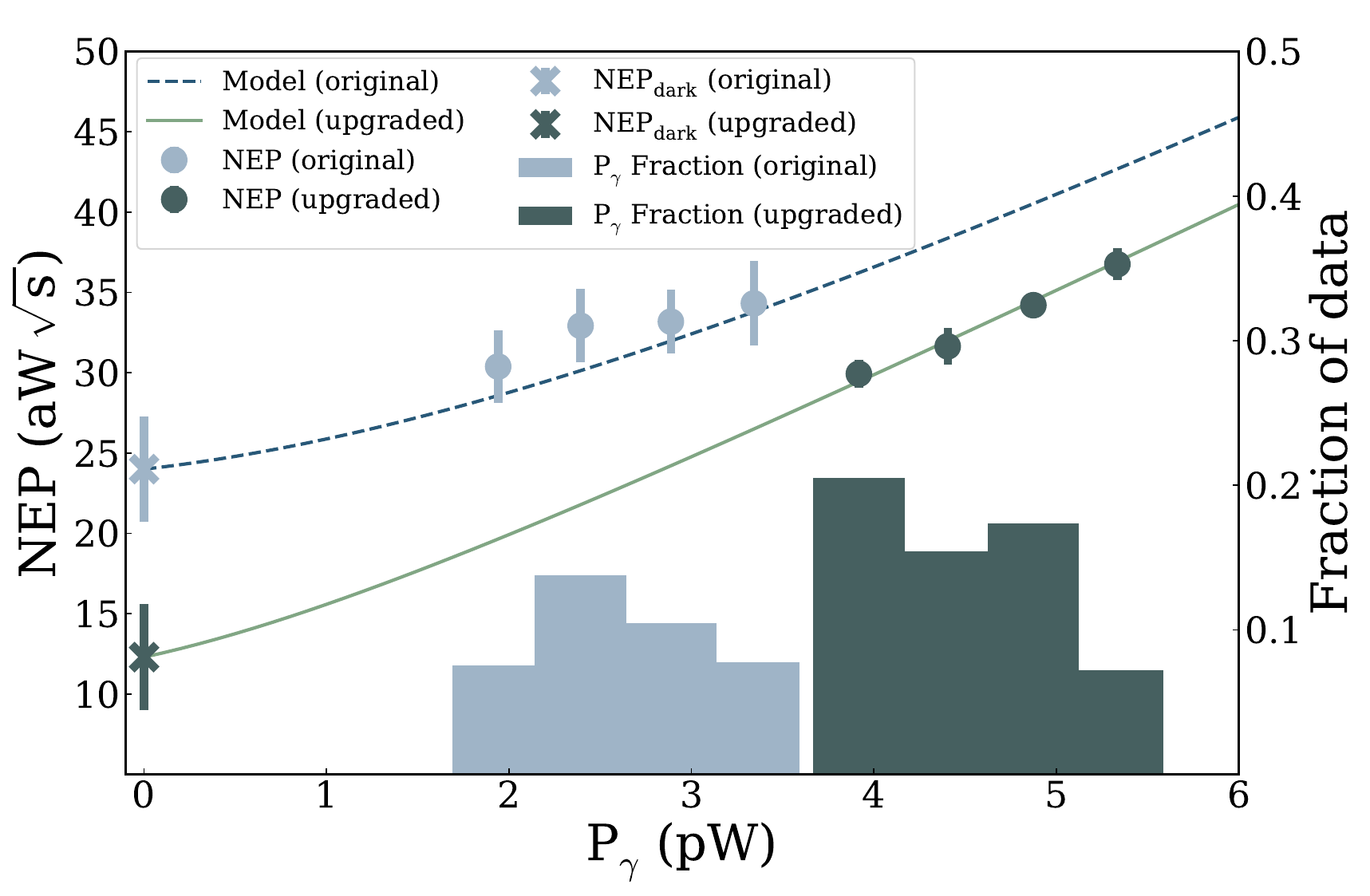}
\caption{Detector NEP versus optical loading. The dark green (light blue) data points and left axis give the measured NEP as a function of loading for the updated (original) detectors. The original detectors comprise only the three best modules saved for the upgraded array (Figure \ref{subfig:W1_in_field}). Similarly-colored histograms and the right axis show the fraction of data, binned in $P_\gamma$, contributing to each data point. Fit to these data, the solid green (blue dashed) line shows the model NEP given in Equation~\ref{eqn:model_NEP} for the updated (original) detectors fit to these data.  The black data point at $P_\gamma=0$ is  $\mathrm{NEP}_\mathrm{dark}$ of the updated detectors from in-lab measurements. Despite having a higher $P_\gamma$,  the new detectors have similar NEP to the originals due to their lower $\mathrm{NEP}_\mathrm{dark}$.}
\label{fig:NEP}
\end{figure}

\begin{figure*}
\centering
\subfloat[]{\includegraphics[height=1.75in]{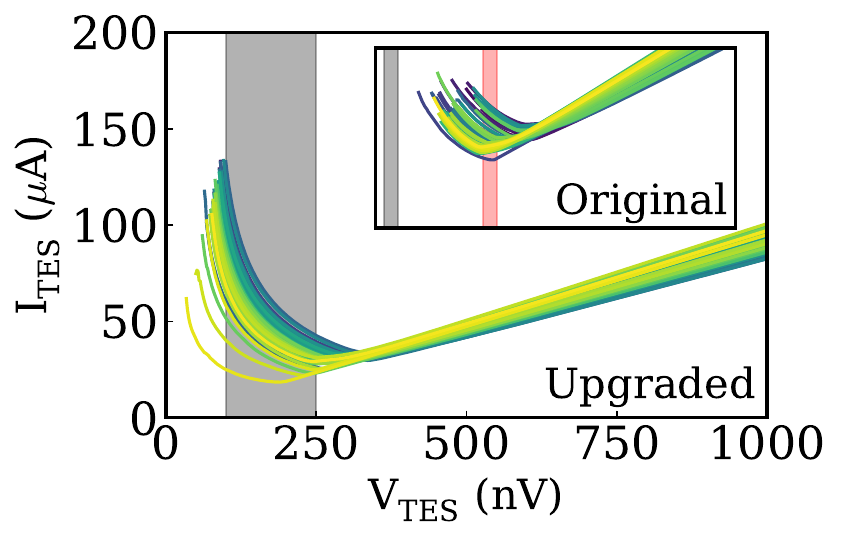}}
\label{subfig:stability_ivs}
\hfil
\subfloat[]{\includegraphics[height=1.8in]{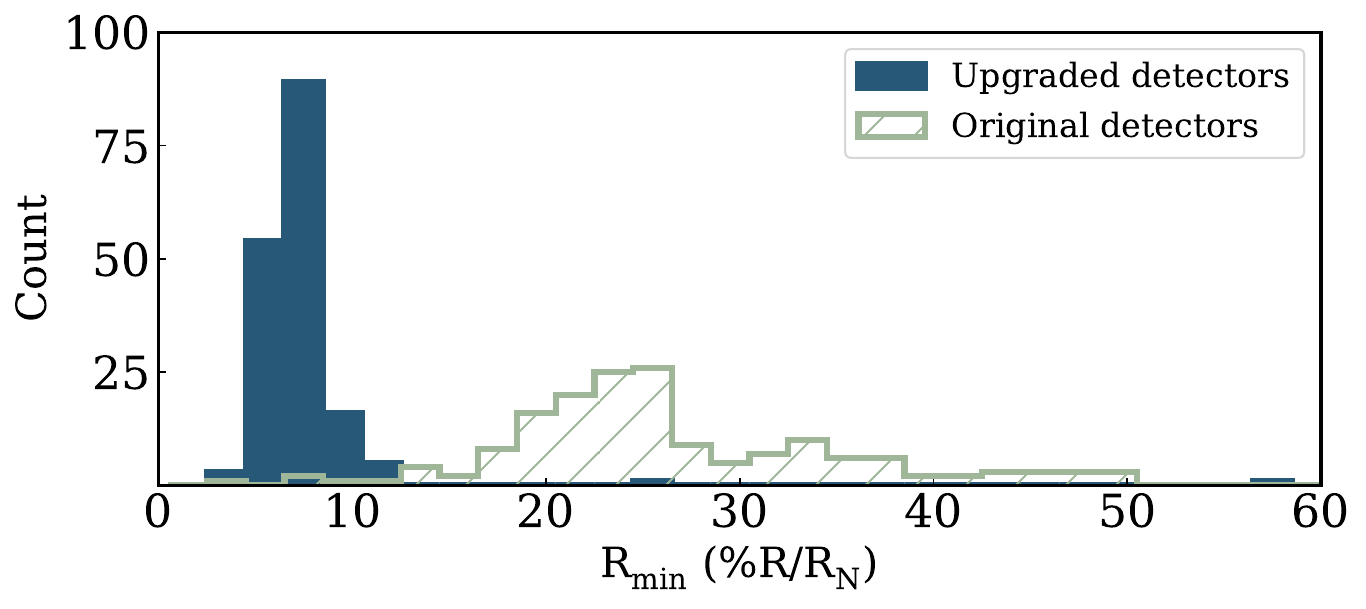}}\label{subfig:percent_Rn}
\caption{$I$--$V$ curves \textbf{(a)} for the upgraded detectors on module 2 show a broad range of bias voltages (shaded region $100$--$250\,\mathrm{nV}$) for which nearly all detectors are on the hyperbola-shaped 
superconducting transition region. The coloring of the curves is only for visual discrimination; it does not correspond to any physical quantity.
The inset plot $I$--$V$ curves from the original detectors on module W33 require a higher bias target (red shaded region). To maintain all detectors biased above the unstable superconducting $I$--$V$ endpoint, a larger proportion of the original detectors must operate near their normal resistance, leading 
to a reduction in the overall array sensitivity.
Figure~\textbf{(b)} shows the minimum percent $R_\mathrm{N}$ below which the detectors become unlocked, in an unstable superconducting state. We find that 94\% of upgraded detectors are stable down to 15\% of their normal resistance. The upgraded detectors allow for a much lower minimum percent $R_\mathrm{N}$, increasing the stability of the detectors and widening the biasing window. }
\label{fig:Rn}
\end{figure*}

CLASS detectors are background-noise limited, i.e., the total $\mathrm{NEP}$ is dominated by the optical loading contribution. The photon NEP can be modeled as:

\begin{equation}
    \left(\mathrm{NEP}_\gamma\right)^2 = h\nu_0P_\gamma + \frac{P^2_\gamma}{\Delta\nu} \,\,\, 
    \, ,
    \label{eqn:model_NEPgamma}
\end{equation}

\noindent where $h$ is Planck's constant. Combining Equations~\ref{eqn:total_NEP} and \ref{eqn:model_NEPgamma}, the total NEP is:
\begin{equation}
    \left(\mathrm{NEP}\right)^2 = \left(\mathrm{NEP}_\mathrm{dark}\right)^2 + h\nu_0P_\gamma + \frac{P^2_\gamma}{\Delta\nu}
    \, .
    \label{eqn:model_NEP}
\end{equation}
We computed the total $\mathrm{NEP}$ using 10-minute chunks of time-ordered data (TOD) from the dataset described in Section \ref{sec:data}. 
We differenced the TODs for each polarization pair within a pixel to reduce any correlated noise.  The per-detector NEP was then computed by averaging the PSD in the side-bands of the \unit{10}{Hz} modulation frequency (between $8.0 \text{--} 9.0 \,\mathrm{Hz}$ and $11.0 \text{--} 12.0\,\mathrm{Hz}$) and dividing by $\sqrt{2}$ (in units of power) to renormalize the pair-differenced noise amplitude to that of a single detector.  The final NEP results are robust across a wide range of NEP variance cut thresholds (Section~\ref{sec:data}). A summary of all NEP estimates is given in Table~\ref{tab:optical-performance}. In this table, we also give the corresponding values for the original CLASS $90\GHz$ detectors as reported in \citetalias{dahal22}.

Figure~\ref{fig:NEP} shows the measured dark and on-sky NEP compared to the model in Equation~\ref{eqn:model_NEP}.  For the model, we used the central frequency $\nu_0$ of a diffuse Rayleigh--Jeans source, which approximates the primary sources of loading (atmosphere and optics), and the simulated FWHP bandwidth $\Delta\nu$   reported in Table \ref{tab:bandpass}. We also fixed the model at $P_\gamma=0$ to the dark NEP value.
The optical loading $P_\gamma$ is estimated as the difference between $P_{\mathrm{sat}}$, measured in laboratory dark tests, and $P_J$ from IV curves at the detector bias point (typically corresponding to $R_{\mathrm{TES}}=30\%R_{\mathrm{N}}$). 
As discussed in \cite{appel19} (Section 2.1), the saturation power of dark detectors--those bolometers in the array not connected to OMT antennas--drops during sky observations compared to lab measurements taken with the array inside a 1 Kelvin can. We assume this $\Psat$ offset impacts both dark and optical detectors equally, leading to an overestimation of on-sky $P_\gamma$ for optical detectors.
To account for this, we introduce another parameter $P_{\mathrm{offset}}$ and fit the model (Equation \ref{eqn:model_NEP}) shifted with the substitution $P_\gamma \rightarrow P_\gamma - P_{\mathrm{offset}}$. We found that $P_{\mathrm{offset}}=1.1\pW$ gives a good fit to the data for both the original and upgraded detectors. As in \cite{appel19}, we hypothesize that the offset is due to optical heating of the detector wafer temperature (and thus increase $\Tbath$ in Equation \ref{eqn:Psat}) above that measured by the focal-plane thermometers. 
Besides $P_{\mathrm{offset}}$, there are no tunable parameters to the model, which can be seen to fit the data in Figure~\ref{fig:NEP} reasonably well. 

The data for ``Original'' wafers included in Figure~\ref{fig:NEP} are from the best three wafers from the original array, which were retained in the upgraded version. (This is why the NEP of the original detectors of $\sim33\,\mathrm{aW\sqrt{s}}$ in Figure~\ref{fig:NEP} is significantly lower than the average of $47\,\mathrm{aW\sqrt{s}}$ quoted for the full original array in Table~\ref{tab:optical-performance}.) 

As discussed in Section~\ref{sec:efficiency}, the upgraded detectors have, on average, higher optical efficiency compared to the originals. This is why, despite the two sets of detectors having similar external (e.g., atmospheric) loading, the upgraded detectors receive a higher optical load in Figure~\ref{fig:NEP}. From Equation~\ref{eqn:model_NEP}, one would expect the higher $P_\gamma$ of the upgraded wafers to result in a higher NEP. However, we see in Figure \ref{fig:NEP} that the NEP is similar between the upgraded detectors and the originals. This is due to the significantly lower $\mathrm{NEP}_\mathrm{dark}$ of the upgraded detectors. This is partially explained by the lower thermal conductivity of the upgraded detectors. Additionally, the original detectors exhibited significant excess noise (\cite{Ullom-zebra}) above the expected phonon noise level. This excess noise is present in the upgraded detectors but at a much lower level.
We conjecture that excess noise in the original detectors stems from internal thermal fluctuations between the TES bilayer and the Pd (\cite{TES_absorber_coupling_2021}). 
In the original TES design the MoAu bilayer is connected to the Pd through long superconducting Nb leads which establish a weak thermal link. 
In the upgraded design, MoAu is linked directly to Pd via a normal metal contact, creating a robust thermal connection that minimizes internal thermal fluctuation noise. This internal thermal noise peaks at kHz frequencies, and is aliased down to the signal band due to the limited bandwidth of the time domain multiplexing scheme of the MCE readout.

With the average NEP for the upgraded wafers taken over the dataset shown in Figure~\ref{fig:NEP}, we estimated the noise equivalent temperature (NET) for the $i^{\mathrm{th}}$ detector, using $\mathrm{NET}_i \approx \mathrm{NEP}_\mathrm{avg} \left(\dfrac{dT_\mathrm{CMB}}{dP_\gamma}\right)_i$.  The calibration factor between optical power deposited on the bolometer to CMB thermodynamic temperature is given by
\begin{equation}
    \dfrac{dT_\mathrm{CMB}}{dP_\gamma} = \dfrac{dT_\mathrm{CMB}}{dT_\mathrm{RJ}} \dfrac{dT_\mathrm{RJ}}{dP_\gamma} \, ,
\end{equation}
where calibration from Rayleigh--Jeans (antenna) temperature to CMB temperature is 
\begin{equation}
    \dfrac{dT_\mathrm{CMB}}{dT_\mathrm{RJ}} = \dfrac{\left( e^{x_0} - 1 \right)^2}{x_0^2 e^{x_0}} \, ,
\end{equation}
with $x_0 = h\nu_0 / k_B T_\mathrm{CMB}$, and the conversion from optical power to Rayleigh--Jeans temperature is 
\begin{equation}
    \dfrac{dT_\mathrm{RJ}}{dP_\gamma} = \dfrac{1}{\eta_i k_B \Delta\nu } \, ,
\end{equation}
with $\eta_i$ being the per-detector absolute efficiency from Section~\ref{sec:efficiency} and $\Delta\nu$ being the simulated FWHP bandwidth from Table~\ref{tab:bandpass}. In Table \ref{tab:optical-performance}, we give the median detector NET and 68\% confidence region for each module.  We inverse-variance weight the per-detector NETs along with those computed for the original detectors as described in \citetalias{dahal22} to obtain  $\mathrm{NET}_\mathrm{array} = 9.7\,\mathrm{\mu K}\,\sqrt{\mathrm{s}}$ for the new array. This is a significant improvement over the original CLASS $90\GHz$ array, which had an array NET of $19\,\mathrm{\mu K}\,\sqrt{\mathrm{s}}$ \citepalias{dahal22}.

\section{Bias and Stability Improvements}
\label{sec:stability}

Each CLASS \unit{90}{GHz} detector module accepts a single bias-current line from the MCE. The single bias-current runs through an array of identical shunt resistors to provide a single-valued voltage bias ($V_\mathrm{TES}$) to all TESs on the detector wafer. We choose $V_\mathrm{TES}$ to maximize the number of detectors operating at stable points on their superconducting transitions. Conversely, detector wafers with TESs with wide and overlapping voltage bias ranges best accommodate the single $V_\mathrm{TES}$ value. By satisfying these conditions, the upgraded detectors improve  the overall bias-ability of the $90\GHz$ focal plane.

The left side of Figure \ref{fig:Rn} shows a family of $I$--$V$ curves from detectors on a single wafer (wafer 2). The data were measured in lab during dark testing at a bath temperature of \unit{75}{mK}.  To set the bias, $V_\mathrm{TES}$ is driven high to force all detectors into their normal state (linear region of the $I$--$V$). It is then swept down in steps, bringing the detector from the normal state onto its superconducting transition (hyperbolic region of the $I$--$V$). The final value of $V_\mathrm{TES}$ is chosen to maximize the number of TESs biased near the centers of their transitions.  As seen in the figure, there is an ample range in $V_\mathrm{TES}$ ($100-250\,\mathrm{nV}$) for which nearly all detectors are on their transitions. 

A useful measure of the quality and width of a detector's superconducting transition is the minimum resistance $R_\mathrm{min}$, expressed as a fraction of $R_\mathrm{N}$, below which a TES falls into the superconducting state. This is also a measure of the ``stability'' of the TES against the onset of electro-thermal runaway at low resistance. The right side of Figure \ref{fig:Rn} shows that $R_\mathrm{min}$ for the upgraded detectors is uniformly low with 94\% of detectors reaching $R_\mathrm{min}<0.15R_\mathrm{N}$. In contrast, the original detectors exhibit a much larger spread and over higher values of $R_\mathrm{min}$. A possible reason for the improvement in stability of the upgraded detectors is the addition of a metallic connection between the MoAu superconducting element and the Pd metallization that holds most of the bolometer's heat capacity.

Finally, while the data in Figure \ref{fig:Rn} are based on dark tests, during observations the TES bias is set while the detector experiences an optical load. This load moves the transition region of the $I$--$V$ curve to lower values. The loading is due primarily to the CMB, the atmosphere, and the optics, which are mostly common to all detectors in a module. Therefore, in principle, optical loading should change the ideal bias voltage in the same way for all detectors, and detectors that shared a common bias in the dark will share a common bias, albeit a different bias, during observations. In practice, however, if the optical efficiencies of the detectors vary too much within a module, the fraction of the loading absorbed will be different for different detectors, creating a variance in the range of acceptable $V_\mathrm{TES}$ among detectors in the same module. Therefore, the improvement in the uniformity of optical efficiencies achieved with the upgraded design (Figures~\ref{fig:histogram_optical_effs} and \ref{fig:efficiencies})  also helps us to maximize the number of simultaneously biased detectors and thus improves effective detector yield.

\section{Conclusions} \label{sec:conclusion}

In this paper, we have reported on the performance of upgraded detectors for the CLASS $90\GHz$ observing band. Four out of seven wafers in the $90\GHz$ detector array were replaced with upgraded versions. Minor changes to the sensor implementation were motivated by lessons learned modeling the 150/220 detector array. For example, the strategy used for termination of the planar OMT’s magic tee and via-less crossovers was revised. These terminations now use a multi-section sub-wavelength absorptive transformer, which is tolerant to metal properties, proximity effect due to superconducting to normal metal contacts, and fabrication misalignment. The most significant design change was at the TES detector, which now benefits from the same lossy transformer approach for the microwave signal absorption, a reduction in effective enclosure volume, and bias choke filters with higher out-of-band rejection. These electromagnetic and detector electro-thermal circuit modifications (e.g., using a contiguous lumped Pd element on the TES membrane to define the detector speed and stability) have significantly improved the observed performance in the deployed detector arrays. Observations of Jupiter confirm that the telescope optical efficiencies are nearly all above 60\%, implying detector efficiencies in excess of 90\%. The detector $I$--$V$ bias curves show that the TESs on a wafer have a broad range of common bias voltages, with 94\% of detectors stable down to 15\% of the normal resistance.
A remaining issue with the detectors is yield in the range of 60\%, which we hypothesize is due to defects on the TES bias leads between each device and the bond pads at the edge of the wafer. Yield can be improved by reducing the length of bias leads (e.g., by routing to multiple edges of the wafer) and by removing design elements and steps that produce defects. In particular, routing the leads to two or four sides of the wafer instead of just one would reduce lead length and densities by one-half to one-fourth, significantly lowering the chances for short or open circuits. This is a straightforward way to increase detector yield for a new CLASS detector module or for future applications of these detectors.

We also report detector bandpass and TES electrothermal parameter measurements that match design values except for a lower-than-expected saturation power. Our measurements suggest that this discrepancy would be explained by the thermal conductance prefactor $\kappa$ being lower than expected. However, we found two alternative scenarios in which $\kappa$ matches expectations: the median $\Tc$ may be $10\mK$ higher than measured or the power-law index for heat conduction may be $n=3.77$ instead of the ideal $n=4$ for ballistic phonon transport. The reality is likely some combination of deviations in $\Tc$, $\kappa$, and $n$ from our baseline scenario. Regardless of these assumptions, the resulting lower detector saturation power together with higher optical efficiency results in detectors with significantly lower noise, reducing the array NET from \unit{19}{\mu K\sqrt{s}} to \unit{9.7}{\mu K\sqrt{s}}. The array sensitivity can be further improved by replacing the remaining three original detector wafers with upgraded versions.

\section*{Acknowledgments}
We acknowledge the National Science Foundation Division of Astronomical Sciences for their support of CLASS under Grant Numbers 0959349, 1429236, 1636634, 1654494, 2034400, and 2109311.
We thank Johns Hopkins University President R. Daniels and the Krieger School of Arts and Sciences Deans for their steadfast support of CLASS.
We further acknowledge the very generous support of Jim Murren and Heather Miller (JHU A\&S '88), Matthew Polk (JHU A\&S Physics BS '71), David Nicholson, and Michael Bloomberg (JHU Engineering '64).
The CLASS project employs detector technology developed in collaboration between JHU and Goddard Space Flight Center under several NASA grants. Detector development work at JHU was funded by NASA cooperative agreement 80NSSC19M0005.
We acknowledge scientific and engineering contributions from Max Abitbol, Fletcher Boone, David Carcamo, Lance Corbett, Ted Grunberg, Saianeesh Haridas, Jake Hassan, Connor Henley, Ben Keller, Lindsay Lowry, Nick Mehrle, Sasha Novak, Diva Parekh, Isu Ravi, Gary Rhodes, Daniel Swartz, Bingjie Wang, Qinan Wang, Tiffany Wei, Zi\'ang Yan, and Zhuo Zhang. We thank Miguel Angel D\'iaz, Joseph Zolenas, Jill Hanson, William Deysher, and Chantal Boisvert for logistical support.
We acknowledge productive collaboration with the JHU Physical Sciences Machine Shop team.
CLASS is located in the Parque Astron\'omico Atacama in northern Chile under the auspices of the Agencia Nacional de Investigaci\'on y Desarrollo (ANID).

\appendix

\section{Detector Property Spatial Variation}
\label{app:Detector Property Spatial Variation}

\begin{figure*}[ht!]
\centering
\includegraphics[width=\textwidth]{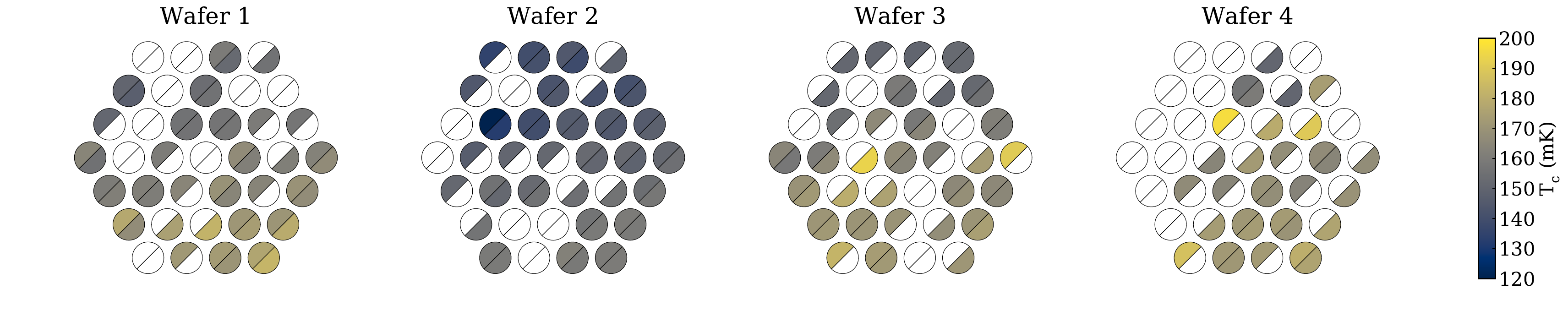}
\includegraphics[width=\textwidth]{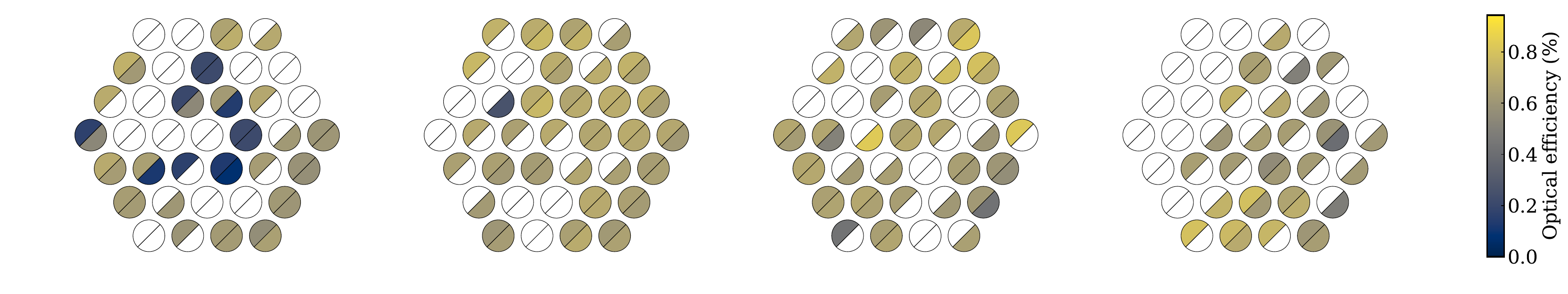}
\includegraphics[width=\textwidth]{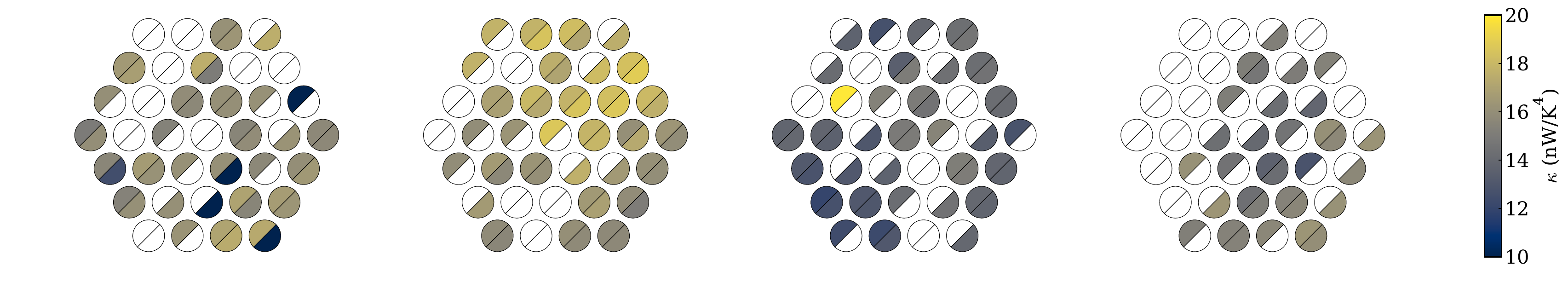}
\includegraphics[width=\textwidth]{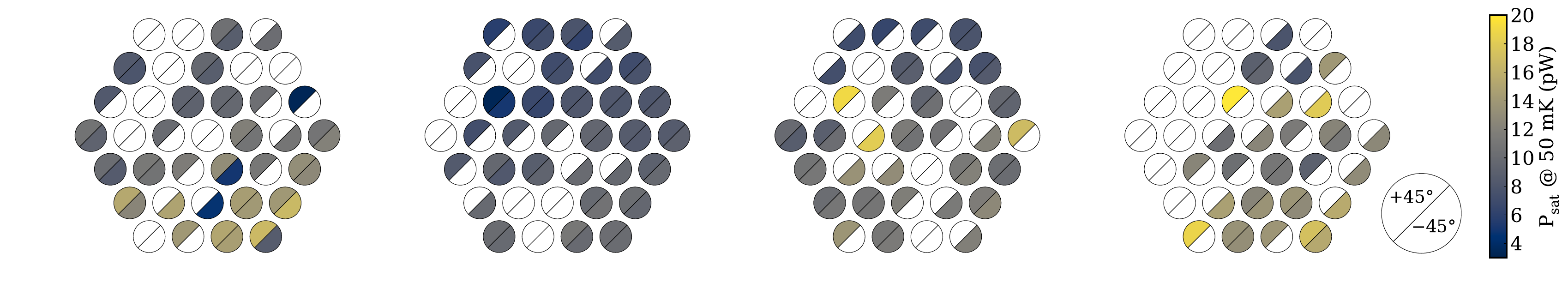}
\caption{The mapping of several electrothermal parameters from Table~\ref{tab:parameters} and optical efficiency onto the 37 pixels of each upgraded wafer.  Each pixel contains two detectors, corresponding to $+45\degree$ (top left of corresponding circles) and $-45\degree$ (bottom right of corresponding circles) as shown in Figure~\ref{subfig:revised_pixel}.  No color indicates an absence of a working detector. Low detector yield is likely due to opens and shorts along the bias leads extending from the detector pixels to the bondpads at the edge of the wafer. 
}
\label{fig:wafer_plots}
\end{figure*}

Figure \ref{fig:wafer_plots} shows the distribution of electrothermal properties ($\Tc$, $\kappa$, and $\Psat$) and telescope optical efficiency ($\eta$) across each of the upgraded wafers. For each pixel on the wafer, the value for each $+45^\circ$ ($-45^\circ$) detector is shown in the top left (bottom right) of the associated circle.  For the most part, the properties are similar between pairs of detectors, and the properties from pixel to pixel vary in a continuous manner (generally linearly) across a wafer. This is what one expects from monotonic variations across the wafer in terms of film depositions, photolithography, and temperature gradients. Wafer~1 shows more spatially random variations than the other wafers. This is likely due to problems encountered during fabrication such that Wafer~1 was designated as ``B grade''. 

The trends in the electrothermal parameters have expected correlations. For instance, because $\Psat \approx \kappa \Tc^n$ we see that the saturation power increases in the same pattern and with non-linear scaling compared to critical temperature. As plotted in Figure \ref{fig:wafer_plots}, the wiring is brought out to bond pads at the bottom of each hexagon of pixels. No trends in the spatial variation of parameters appear to be associated with this symmetry-breaking feature. For the ``A grade'' Wafers 2, 3, and 4, the telescope optical efficiency is nearly uniform across the wafer and even from wafer to wafer. This indicates that the microwave circuit design is robust to modest fabrication variations that impact the TES electrothermal parameters. Because the optical efficiency includes the effects of the whole telescope, the uniformity of the efficiency also indicates that the optical performance (e.g., beam spill) of the telescope is uniform across the focal plane. 
 
\section{Parameter Quick Reference}
\label{app:parameter index}
\begin{deluxetable*}{cccc}

 The detector parameters defined in the main text are referenced in Table~\ref{tab:parameter-index}, along with their estimation method, equations, and an index of sections where each appears.

\tablecaption{Quick Reference and Section Index for Parameters\label{tab:parameter-index}}

\tablehead{Name & Symbol  & Estimation Method and Equations & Sections\tablenotemark{a} }
 \renewcommand{\arraystretch}{1.5}
\startdata
Bias (Joule) Power & $P_J$ & Measured $P_J=I_{\mathrm{TES}}\times V_{\mathrm{TES}}$ & \ref{subsec:Rn and Psat}$^*$, \ref{subsec:tau and C} \\\hline
Saturation Power & $P_{\mathrm{sat}}$ & $\Psat=P_J$ for $P_\gamma=0$ and $R_{TES}=0.7R_N$ & \ref{subsec:Rn and Psat}$^*$ \\\hline
Photon Power & $P_\gamma$ & $P_\gamma=P_{\mathrm{sat}}-P_J$ &\ref{subsec:tau and C}$^*$, \ref{sec:noise} \\\hline
Critical Temperature & $T_c$ & $P_{\mathrm{sat}} = \kappa \left(T_\mathrm{c}^n - T_{\mathrm{bath}}^n\right)$ fit to IV Data & \ref{subsec:TES Thermal Circuit}, \ref{subsec:Tc and G}$^*$, \ref{ssec:Alternative Scenarios}, \ref{sec:noise} \\\hline
Bath Temperature & $T_{\mathrm{bath}}$ & Measured by Thermometry & \ref{subsec:TES Thermal Circuit}, \ref{subsec:Tc and G}, \ref{sec:noise} \\\hline
Normal Resistance & $R_\mathrm{N}$ & $I_\mathrm{TES}=V_\mathrm{TES}/R_N$ fit to normal-TES IV data & \ref{subsec:Rn and Psat}$^*$ \\\hline
Thermal Conductance Prefactor & $\kappa$ & $P_{\mathrm{sat}} = \kappa \left(T_\mathrm{c}^n - T_{\mathrm{bath}}^n\right)$ fit to IV Data & \ref{subsec:Tc and G}$^*$, \ref{ssec:Alternative Scenarios} \\\hline
Thermal Conductance Index & $n$ & $n=4$ for ballistic phonons & \ref{subsec:Tc and G}, \ref{ssec:Alternative Scenarios} \\\hline
Thermal Conductance\tablenotemark{b} & $G$ & $G=n \kappa T_c^{n-1} \approx n\Psat/\Tc$ & \ref{subsec:Tc and G}$^*$\\\hline
Heat Capacity & $C$ & $C=G\tau$ & \ref{subsec:TES Thermal Circuit}, \ref{subsec:tau and C}$^*$, \ref{ssec:Alternative Scenarios} \\\hline
Thermal Time Constant & $\tau$ & \cite[][Eq. 26]{appel22} given $\tau_\gamma$ & \ref{subsec:tau and C}$^*$ \\\hline
Optical Time Constant & $\tau_\gamma$ & Measured from VPM Emission & \ref{subsec:tau and C}$^*$ \\\hline
Bandwidth &$\Delta \nu$&  Simulation checked with FTS measurement &\ref{sec:bandpass}$^*$\\\hline
Bandcenter &$\nu_0$& Simulation checked with FTS measurement &\ref{sec:bandpass}$^*$\\\hline
Telescope Optical Efficiency& $\eta$ & Measured with Planet Observations & \ref{sec:efficiency}$^*$ \\\hline
Phonon NEP & $\mathrm{NEP}_{\mathrm{phonon}}$ & $\mathrm{NEP}_\mathrm{phonon} = (2 k_B T_\mathrm{c}^2 G F_\mathrm{link})^{1/2}$ & \ref{sec:noise} \\ \hline
 & $F_{\mathrm{link}}$ & $F_{\mathrm{link, phonon}}=(1+ (T_\mathrm{bath}/T_\mathrm{c})^{n+1})/2\approx 1/2$ & \ref{sec:noise} \\\hline
Dark NEP\tablenotemark{c} &  $\mathrm{NEP}_\mathrm{dark}$  &Measured in Dark Configuration & \ref{sec:noise}$^*$ \\\hline
Photon NEP & $\mathrm{NEP}_\gamma$ & $\left(\mathrm{NEP}_\gamma\right)^2 = h\nu_0 P_\gamma + P^2_\gamma/\Delta\nu$ & \ref{sec:noise}$^*$\\\hline
\multirow{2}{*}{Total NEP} & \multirow{2}{*}{$\mathrm{NEP}$} & Measured from Survey Data & 
\multirow{2}{*}{\ref{sec:noise}$^*$} \\
& & $\left(\mathrm{NEP}\right)^2 = \left(\mathrm{NEP}_\gamma\right)^2 + \left(\mathrm{NEP}_\mathrm{dark}\right)^2$ & \\\hline
\enddata
\tablenotetext{a}{Sections with a star denote where measurements are described. Otherwise section designation indicates parts of the text in which the parameter is described or plays a role in the discussion.}
\tablenotetext{b}{The approximation given for this parameter assumes $T_c^n\gg T_{\mathrm{bath}}^n$, which is valid for CLASS. }
\tablenotetext{c}{Dark NEP includes the Phonon NEP as well as any other contributions to total NEP (e.g., from the SQUID readout or excess noise due to a non-trivial TES thermal circuit) besides the photon NEP.}
\end{deluxetable*}

\bibliography{main,cmb,class_pub,hardware}{}

\begin{thebibliography}{}
\expandafter\ifx\csname natexlab\endcsname\relax\def\natexlab#1{#1}\fi
\providecommand{\url}[1]{\href{#1}{#1}}
\providecommand{\dodoi}[1]{doi:~\href{http://doi.org/#1}{\nolinkurl{#1}}}
\providecommand{\doeprint}[1]{\href{http://ascl.net/#1}{\nolinkurl{http://ascl.net/#1}}}
\providecommand{\doarXiv}[1]{\href{https://arxiv.org/abs/#1}{\nolinkurl{https://arxiv.org/abs/#1}}}

\bibitem[{{Addamo} {et~al.}(2021){Addamo}, {Ade}, {Baccigalupi}, {Baldini},
  {Battaglia}, {Battistelli}, {Ba{\`u}}, {de Bernardis}, {Bersanelli},
  {Biasotti}, {Boscaleri}, {Caccianiga}, {Caprioli}, {Cavaliere}, {Cei},
  {Cleary}, {Columbro}, {Coppi}, {Coppolecchia}, {Cuttaia}, {D'Alessandro}, {De
  Gasperis}, {De Petris}, {Fafone}, {Farsian}, {Ferrari Barusso}, {Fontanelli},
  {Franceschet}, {Gaier}, {Galli}, {Gatti}, {Genova-Santos}, {Gerbino},
  {Gervasi}, {Ghigna}, {Grosso}, {Gruppuso}, {Gualtieri}, {Incardona}, {Jones},
  {Kangaslahti}, {Krachmalnicoff}, {Lamagna}, {Lattanzi},
  {L{\'o}pez-Caraballo}, {Lumia}, {Mainini}, {Maino}, {Mandelli}, {Maris},
  {Masi}, {Matarrese}, {May}, {Mele}, {Mena}, {Mennella}, {Molina}, {Molinari},
  {Morgante}, {Natale}, {Nati}, {Natoli}, {Pagano}, {Paiella}, {Panico},
  {Paonessa}, {Paradiso}, {Passerini}, {Perez-de-Taoro}, {Peverini},
  {Pezzotta}, {Piacentini}, {Piccirillo}, {Pisano}, {Polenta}, {Poletti},
  {Presta}, {Realini}, {Reyes}, {Rocchi}, {Rubino-Martin}, {Sandri}, {Sartor},
  {Schillaci}, {Signorelli}, {Siri}, {Soria}, {Spinella}, {Tapia}, {Tartari},
  {Taylor}, {Terenzi}, {Tomasi}, {Tommasi}, {Tucker}, {Vaccaro}, {Vigano},
  {Villa}, {Virone}, {Vittorio}, {Volpe}, {Watkins}, {Zacchei}, {Zannoni}, \&
  {LSPE Collaboration}}]{lspe20}
{Addamo}, G., {Ade}, P.~A.~R., {Baccigalupi}, C., {et~al.} 2021, \jcap, 2021,
  008, \dodoi{10.1088/1475-7516/2021/08/008}

\bibitem[{{Ali} {et~al.}(2022){Ali}, {Essinger-Hileman}, {Marriage}, {Appel},
  {Bennett}, {Berkeley}, {Bulcha}, {Chuss}, {Dahal}, {Denis}, {Rostem},
  {U-Yen}, {Wollack}, \& {Zeng}}]{Ali-CE7}
{Ali}, A.~M., {Essinger-Hileman}, T., {Marriage}, T., {et~al.} 2022, Review of
  Scientific Instruments, 93, 024503, \dodoi{10.1063/5.0049526}

\bibitem[{{Appel} {et~al.}(2014){Appel}, {Ali}, {Amiri}, {Araujo}, {Bennet},
  {Boone}, {Chan}, {Cho}, {Chuss}, {Colazo}, {Crowe}, {Denis}, {D{\"u}nner},
  {Eimer}, {Essinger-Hileman}, {Gothe}, {Halpern}, {Harrington}, {Hilton},
  {Hinshaw}, {Huang}, {Irwin}, {Jones}, {Karakula}, {Kogut}, {Larson}, {Limon},
  {Lowry}, {Marriage}, {Mehrle}, {Miller}, {Miller}, {Moseley}, {Novak},
  {Reintsema}, {Rostem}, {Stevenson}, {Towner}, {U-Yen}, {Wagner}, {Watts},
  {Wollack}, {Xu}, \& {Zeng}}]{appel14spie}
{Appel}, J.~W., {Ali}, A., {Amiri}, M., {et~al.} 2014, in Society of
  Photo-Optical Instrumentation Engineers (SPIE) Conference Series, Vol. 9153,
  Millimeter, Submillimeter, and Far-Infrared Detectors and Instrumentation for
  Astronomy VII, 91531J, \dodoi{10.1117/12.2056530}

\bibitem[{{Appel} {et~al.}(2019){Appel}, {Xu}, {Padilla}, {Harrington},
  {Pradenas Marquez}, {Ali}, {Bennett}, {Brewer}, {Bustos}, {Chan}, {Chuss},
  {Cleary}, {Couto}, {Dahal}, {Denis}, {D{\"u}nner}, {Eimer},
  {Essinger-Hileman}, {Fluxa}, {Gothe}, {Hilton}, {Hubmayr}, {Iuliano},
  {Karakla}, {Marriage}, {Miller}, {N{\'u}{\~n}ez}, {Parker}, {Petroff},
  {Reintsema}, {Rostem}, {Stevens}, {Nunes Valle}, {Wang}, {Watts}, {Wollack},
  \& {Zeng}}]{appel19}
{Appel}, J.~W., {Xu}, Z., {Padilla}, I.~L., {et~al.} 2019, \apj, 876, 126,
  \dodoi{10.3847/1538-4357/ab1652}

\bibitem[{{Appel} {et~al.}(2022){Appel}, {Bennett}, {Brewer}, {Bustos}, {Chan},
  {Chuss}, {Cleary}, {Couto}, {Dahal}, {Datta}, {Denis}, {Eimer},
  {Essinger-Hileman}, {Harrington}, {Iuliano}, {Li}, {Marriage},
  {N{\'u}{\~n}ez}, {Osumi}, {Padilla}, {Petroff}, {Rostem}, {Valle}, {Watts},
  {Weiland}, {Wollack}, \& {Xu}}]{appel22}
{Appel}, J.~W., {Bennett}, C.~L., {Brewer}, M.~K., {et~al.} 2022, \apjs, 262,
  52, \dodoi{10.3847/1538-4365/ac8cf2}

\bibitem[{{Arnold} {et~al.}(2012){Arnold}, {Ade}, {Anthony}, {Barron},
  {Boettger}, {Borrill}, {Chapman}, {Chinone}, {Dobbs}, {Errard}, {Fabbian},
  {Flanigan}, {Fuller}, {Ghribi}, {Grainger}, {Halverson}, {Hasegawa},
  {Hattori}, {Hazumi}, {Holzapfel}, {Howard}, {Hyland}, {Jaffe}, {Keating},
  {Kermish}, {Kisner}, {Le Jeune}, {Lee}, {Linder}, {Lungu}, {Matsuda},
  {Matsumura}, {Miller}, {Meng}, {Morii}, {Moyerman}, {Myers}, {Nishino},
  {Paar}, {Quealy}, {Reichardt}, {Richards}, {Ross}, {Shimizu}, {Shimmin},
  {Shimon}, {Sholl}, {Siritanasak}, {Spieler}, {Stebor}, {Steinbach},
  {Stompor}, {Suzuki}, {Tomaru}, {Tucker}, \& {Zahn}}]{arnold12}
{Arnold}, K., {Ade}, P.~A.~R., {Anthony}, A.~E., {et~al.} 2012, in Society of
  Photo-Optical Instrumentation Engineers (SPIE) Conference Series, Vol. 8452,
  Millimeter, Submillimeter, and Far-Infrared Detectors and Instrumentation for
  Astronomy VI, ed. W.~S. {Holland} \& J.~{Zmuidzinas}, 84521D,
  \dodoi{10.1117/12.927057}

\bibitem[{{Battistelli} {et~al.}(2008){Battistelli}, {Amiri}, {Burger},
  {Halpern}, {Knotek}, {Ellis}, {Gao}, {Kelly}, {Macintosh}, {Irwin}, \&
  {Reintsema}}]{Battistelli08}
{Battistelli}, E.~S., {Amiri}, M., {Burger}, B., {et~al.} 2008, Journal of Low
  Temperature Physics, 151, 908, \dodoi{10.1007/s10909-008-9772-z}

\bibitem[{{Bennett} {et~al.}(2013){Bennett}, {Larson}, {Weiland}, {Jarosik},
  {Hinshaw}, {Odegard}, {Smith}, {Hill}, {Gold}, {Halpern}, {Komatsu}, {Nolta},
  {Page}, {Spergel}, {Wollack}, {Dunkley}, {Kogut}, {Limon}, {Meyer}, {Tucker},
  \& {Wright}}]{bennett13}
{Bennett}, C.~L., {Larson}, D., {Weiland}, J.~L., {et~al.} 2013, \apjs, 208,
  20, \dodoi{10.1088/0067-0049/208/2/20}

\bibitem[{{BICEP/Keck Collaboration} {et~al.}(2021){BICEP/Keck Collaboration},
  {Ade}, {Ahmed}, {Amiri}, {Barkats}, {Thakur}, {Bischoff}, {Beck}, {Bock},
  {Boenish}, {Bullock}, {Buza}, {Cheshire}, {Connors}, {Cornelison},
  {Crumrine}, {Cukierman}, {Denison}, {Dierickx}, {Duband}, {Eiben},
  {Fatigoni}, {Filippini}, {Fliescher}, {Goeckner-Wald}, {Goldfinger},
  {Grayson}, {Grimes}, {Hall}, {Halal}, {Halpern}, {Hand}, {Harrison},
  {Henderson}, {Hildebrandt}, {Hilton}, {Hubmayr}, {Hui}, {Irwin}, {Kang},
  {Karkare}, {Karpel}, {Kefeli}, {Kernasovskiy}, {Kovac}, {Kuo}, {Lau},
  {Leitch}, {Lennox}, {Megerian}, {Minutolo}, {Moncelsi}, {Nakato}, {Namikawa},
  {Nguyen}, {O'Brient}, {Ogburn}, {Palladino}, {Prouve}, {Pryke}, {Racine},
  {Reintsema}, {Richter}, {Schillaci}, {Schwarz}, {Schmitt}, {Sheehy},
  {Soliman}, {Germaine}, {Steinbach}, {Sudiwala}, {Teply}, {Thompson}, {Tolan},
  {Tucker}, {Turner}, {Umilt{\`a}}, {Verg{\`e}s}, {Vieregg}, {Wandui}, {Weber},
  {Wiebe}, {Willmert}, {Wong}, {Wu}, {Yang}, {Yoon}, {Young}, {Yu}, {Zeng},
  {Zhang}, \& {Zhang}}]{BK21}
{BICEP/Keck Collaboration}, {Ade}, P.~A.~R., {Ahmed}, Z., {et~al.} 2021, \prl,
  127, 151301, \dodoi{10.1103/PhysRevLett.127.151301}

\bibitem[{{Boyle} \& {Rodgers}(1959)}]{Boyle}
{Boyle}, W.~S., \& {Rodgers}, K.~F., J. 1959, Journal of the Optical Society of
  America (1917-1983), 49, 66

\bibitem[{{Chuss} {et~al.}(2016){Chuss}, {Ali}, {Amiri}, {Appel}, {Bennett},
  {Colazo}, {Denis}, {D{\"u}nner}, {Essinger-Hileman}, {Eimer}, {Fluxa},
  {Gothe}, {Halpern}, {Harrington}, {Hilton}, {Hinshaw}, {Hubmayr}, {Iuliano},
  {Marriage}, {Miller}, {Moseley}, {Mumby}, {Petroff}, {Reintsema}, {Rostem},
  {U-Yen}, {Watts}, {Wagner}, {Wollack}, {Xu}, \& {Zeng}}]{Chuss-development}
{Chuss}, D.~T., {Ali}, A., {Amiri}, M., {et~al.} 2016, Journal of Low
  Temperature Physics, 184, 759, \dodoi{10.1007/s10909-015-1368-9}

\bibitem[{{Cleary}(2023)}]{clearyphdthesis}
{Cleary}, J. 2023, PhD thesis, Johns Hopkins University

\bibitem[{{CMB-S4 Collaboration} {et~al.}(2022){CMB-S4 Collaboration},
  {Abazajian}, {Addison}, {Adshead}, {Ahmed}, {Akerib}, {Ali}, {Allen},
  {Alonso}, {Alvarez}, {Amin}, {Anderson}, {Arnold}, {Ashton}, {Baccigalupi},
  {Bard}, {Barkats}, {Barron}, {Barry}, {Bartlett}, {Basu Thakur}, {Battaglia},
  {Bean}, {Bebek}, {Bender}, {Benson}, {Bianchini}, {Bischoff}, {Bleem},
  {Bock}, {Bocquet}, {Boddy}, {Richard Bond}, {Borrill}, {Bouchet},
  {Brinckmann}, {Brown}, {Bryan}, {Buza}, {Byrum}, {Hervias Caimapo},
  {Calabrese}, {Calafut}, {Caldwell}, {Carlstrom}, {Carron}, {Cecil},
  {Challinor}, {Chang}, {Chinone}, {Sherry Cho}, {Cooray}, {Coulton},
  {Crawford}, {Crites}, {Cukierman}, {Cyr-Racine}, {de Haan}, {Delabrouille},
  {Devlin}, {Di Valentino}, {Dierickx}, {Dobbs}, {Duff}, {Dvorkin}, {Eimer},
  {Elleflot}, {Errard}, {Essinger-Hileman}, {Fabbian}, {Feng}, {Ferraro},
  {Filippini}, {Flauger}, {Flaugher}, {Fraisse}, {Frolov}, {Galitzki},
  {Gallardo}, {Galli}, {Ganga}, {Gerbino}, {Gluscevic}, {Goeckner-Wald},
  {Green}, {Grin}, {Grohs}, {Gualtieri}, {Gudmundsson}, {Gullett}, {Gupta},
  {Habib}, {Halpern}, {Halverson}, {Hanany}, {Harrington}, {Hasegawa},
  {Hasselfield}, {Hazumi}, {Heitmann}, {Henderson}, {Hensley}, {Hill}, {Colin
  Hill}, {Hlo{\v{z}}ek}, {Patty Ho}, {Hoang}, {Holder}, {Holzapfel}, {Hood},
  {Hubmayr}, {Huffenberger}, {Hui}, {Irwin}, {Jeong}, {Johnson}, {Jones}, {Hwan
  Kang}, {Karkare}, {Katayama}, {Keskitalo}, {Kisner}, {Knox}, {Koopman},
  {Kosowsky}, {Kovac}, {Kovetz}, {Kuhlmann}, {Kuo}, {Kusaka},
  {L{\"a}hteenm{\"a}ki}, {Lawrence}, {Lee}, {Lewis}, {Li}, {Linder}, {Loverde},
  {Lowitz}, {Lubin}, {Madhavacheril}, {Mantz}, {Marques}, {Matsuda},
  {Mauskopf}, {McCarrick}, {McMahon}, {Daniel Meerburg}, {Melin}, {Menanteau},
  {Meyers}, {Millea}, {Mohr}, {Moncelsi}, {Monzani}, {Mroczkowski},
  {Mukherjee}, {Nagy}, {Namikawa}, {Nati}, {Natoli}, {Newburgh}, {Niemack},
  {Nishino}, {Nord}, {Novosad}, {O'Brient}, {Padin}, {Palladino}, {Partridge},
  {Petravick}, {Pierpaoli}, {Pogosian}, {Prabhu}, {Pryke}, {Puglisi}, {Racine},
  {Rahlin}, {Sathyanarayana Rao}, {Raveri}, {Reichardt}, {Remazeilles},
  {Rocha}, {Roe}, {Roy}, {Ruhl}, {Salatino}, {Saliwanchik}, {Schaan},
  {Schillaci}, {Schmitt}, {Schmittfull}, {Scott}, {Sehgal}, {Shandera},
  {Sherwin}, {Shirokoff}, {Simon}, {Slosar}, {Spergel}, {St. Germaine},
  {Staggs}, {Stark}, {Starkman}, {Stompor}, {Stoughton}, {Suzuki}, {Tajima},
  {Teply}, {Thompson}, {Thorne}, {Timbie}, {Tomasi}, {Tristram}, {Tucker},
  {Umilt{\`a}}, {van Engelen}, {Vavagiakis}, {Vieira}, {Vieregg}, {Wagoner},
  {Wallisch}, {Wang}, {Watson}, {Westbrook}, {Whitehorn}, {Wollack}, {Kimmy
  Wu}, {Xu}, {Eric Yang}, {Yasini}, {Yefremenko}, {Won Yoon}, {Young}, {Yu}, \&
  {Zonca}}]{stagefour22bmodeforecast}
{CMB-S4 Collaboration}, {Abazajian}, K., {Addison}, G.~E., {et~al.} 2022, \apj,
  926, 54, \dodoi{10.3847/1538-4357/ac1596}

\bibitem[{{Crowe} {et~al.}(2013){Crowe}, {Bennett}, {Chuss}, {Denis}, {Eimer},
  {Lourie}, {Marriage}, {Moseley}, {Rostem}, {Stevenson}, {Towner}, {U-yen}, \&
  {Wollack}}]{Crowe-choke}
{Crowe}, E.~J., {Bennett}, C.~L., {Chuss}, D.~T., {et~al.} 2013, IEEE
  Transactions on Applied Superconductivity, 23, 2500505,
  \dodoi{10.1109/TASC.2012.2237211}

\bibitem[{{Dahal} {et~al.}(2018){Dahal}, {Ali}, {Appel}, {Essinger-Hileman},
  {Bennett}, {Brewer}, {Bustos}, {Chan}, {Chuss}, {Cleary}, {Colazo}, {Couto},
  {Denis}, {D{\"u}nner}, {Eimer}, {Engelhoven}, {Fluxa}, {Halpern},
  {Harrington}, {Helson}, {Hilton}, {Hinshaw}, {Hubmayr}, {Iuliano}, {Karakla},
  {Marriage}, {McMahon}, {Miller}, {Nu{\~n}ez}, {Padilla}, {Palma}, {Parker},
  {Petroff}, {Pradenas}, {Reeves}, {Reintsema}, {Rostem}, {Sagliocca}, {U-Yen},
  {Valle}, {Wang}, {Wang}, {Watts}, {Weiland}, {Wollack}, {Xu}, {Yan}, \&
  {Zeng}}]{dahal18spie}
{Dahal}, S., {Ali}, A., {Appel}, J.~W., {et~al.} 2018, in Society of
  Photo-Optical Instrumentation Engineers (SPIE) Conference Series, Vol. 10708,
  Society of Photo-Optical Instrumentation Engineers (SPIE) Conference Series,
  107081Y, \dodoi{10.1117/12.2311812}

\bibitem[{{Dahal} {et~al.}(2020){Dahal}, {Amiri}, {Appel}, {Bennett},
  {Corbett}, {Datta}, {Denis}, {Essinger-Hileman}, {Halpern}, {Helson},
  {Hilton}, {Hubmayr}, {Keller}, {Marriage}, {Nunez}, {Petroff}, {Reintsema},
  {Rostem}, {U-Yen}, \& {Wollack}}]{dahal20HF}
{Dahal}, S., {Amiri}, M., {Appel}, J.~W., {et~al.} 2020, Journal of Low
  Temperature Physics, 199, 289, \dodoi{10.1007/s10909-019-02317-0}

\bibitem[{{Dahal} {et~al.}(2022){Dahal}, {Appel}, {Datta}, {Brewer}, {Ali},
  {Bennett}, {Bustos}, {Chan}, {Chuss}, {Cleary}, {Couto}, {Denis},
  {D{\"u}nner}, {Eimer}, {Espinoza}, {Essinger-Hileman}, {Golec}, {Harrington},
  {Helson}, {Iuliano}, {Karakla}, {Li}, {Marriage}, {McMahon}, {Miller},
  {Novack}, {N{\'u}{\~n}ez}, {Osumi}, {Padilla}, {Palma}, {Parker}, {Petroff},
  {Reeves}, {Rhoades}, {Rostem}, {Valle}, {Watts}, {Weiland}, {Wollack}, \&
  {Xu}}]{dahal22}
{Dahal}, S., {Appel}, J.~W., {Datta}, R., {et~al.} 2022, \apj, 926, 33,
  \dodoi{10.3847/1538-4357/ac397c}

\bibitem[{{Datta} {et~al.}(2024){Datta}, {Brewer}, {Couto}, {Eimer}, {Li},
  {Xu}, {Ali}, {Appel}, {Bennett}, {Bustos}, {Chuss}, {Cleary}, {Dahal}, {Raul
  Javier Espinoza Inostroza}, {Essinger-Hileman}, {Flux{\'a}}, {Harrington},
  {Helson}, {Iuliano}, {Karakla}, {Marriage}, {Novack}, {N{\'u}{\~n}ez},
  {Padilla}, {Parker}, {Petroff}, {Reeves}, {Rostem}, {Shi}, {Valle}, {Watts},
  {Weiland}, {Wollack}, \& {Zeng}}]{datta23}
{Datta}, R., {Brewer}, M.~K., {Couto}, J.~D., {et~al.} 2024, \apjs, 273, 26,
  \dodoi{10.3847/1538-4365/ad50a0}

\bibitem[{{de Wit} {et~al.}(2021){de Wit}, {Gottardi}, {Taralli}, {Nagayoshi},
  {Ridder}, {Akamatsu}, {Bruijn}, {Hoogeveen}, {van der Kuur}, {Ravensberg},
  {Vaccaro}, {Gao}, \& {den Herder}}]{TES_absorber_coupling_2021}
{de Wit}, M., {Gottardi}, L., {Taralli}, E., {et~al.} 2021, Physical Review
  Applied, 16, 044059, \dodoi{10.1103/PhysRevApplied.16.044059}

\bibitem[{{Denis} {et~al.}(2009){Denis}, {Cao}, {Chuss}, {Eimer}, {Hinderks},
  {Hsieh}, {Moseley}, {Stevenson}, {Talley}, {U. -yen}, \&
  {Wollack}}]{Denis-fabrication}
{Denis}, K.~L., {Cao}, N.~T., {Chuss}, D.~T., {et~al.} 2009, in American
  Institute of Physics Conference Series, Vol. 1185, The Thirteenth
  International Workshop on Low Temperature Detectors - LTD13, ed. B.~{Young},
  B.~{Cabrera}, \& A.~{Miller}, 371--374, \dodoi{10.1063/1.3292355}

\bibitem[{{Doriese} {et~al.}(2016){Doriese}, {Morgan}, {Bennett}, {Denison},
  {Fitzgerald}, {Fowler}, {Gard}, {Hays-Wehle}, {Hilton}, {Irwin}, {Joe},
  {Mates}, {O'Neil}, {Reintsema}, {Robbins}, {Schmidt}, {Swetz}, {Tatsuno},
  {Vale}, \& {Ullom}}]{nist_tdm_mux13b}
{Doriese}, W.~B., {Morgan}, K.~M., {Bennett}, D.~A., {et~al.} 2016, Journal of
  Low Temperature Physics, 184, 389, \dodoi{10.1007/s10909-015-1373-z}

\bibitem[{{Eimer} {et~al.}(2012){Eimer}, {Bennett}, {Chuss}, {Marriage},
  {Wollack}, \& {Zeng}}]{eimer12spie}
{Eimer}, J.~R., {Bennett}, C.~L., {Chuss}, D.~T., {et~al.} 2012, in Society of
  Photo-Optical Instrumentation Engineers (SPIE) Conference Series, Vol. 8452,
  Millimeter, Submillimeter, and Far-Infrared Detectors and Instrumentation for
  Astronomy VI, ed. W.~S. {Holland} \& J.~{Zmuidzinas}, 845220,
  \dodoi{10.1117/12.925464}

\bibitem[{{Essinger-Hileman} {et~al.}(2014){Essinger-Hileman}, {Ali}, {Amiri},
  {Appel}, {Araujo}, {Bennett}, {Boone}, {Chan}, {Cho}, {Chuss}, {Colazo},
  {Crowe}, {Denis}, {D{\"u}nner}, {Eimer}, {Gothe}, {Halpern}, {Harrington},
  {Hilton}, {Hinshaw}, {Huang}, {Irwin}, {Jones}, {Karakla}, {Kogut}, {Larson},
  {Limon}, {Lowry}, {Marriage}, {Mehrle}, {Miller}, {Miller}, {Moseley},
  {Novak}, {Reintsema}, {Rostem}, {Stevenson}, {Towner}, {U-Yen}, {Wagner},
  {Watts}, {Wollack}, {Xu}, \& {Zeng}}]{essinger-hileman14spie}
{Essinger-Hileman}, T., {Ali}, A., {Amiri}, M., {et~al.} 2014, in Society of
  Photo-Optical Instrumentation Engineers (SPIE) Conference Series, Vol. 9153,
  Millimeter, Submillimeter, and Far-Infrared Detectors and Instrumentation for
  Astronomy VII, ed. W.~S. {Holland} \& J.~{Zmuidzinas}, 91531I,
  \dodoi{10.1117/12.2056701}

\bibitem[{{Harrington} {et~al.}(2016){Harrington}, {Marriage}, {Ali}, {Appel},
  {Bennett}, {Boone}, {Brewer}, {Chan}, {Chuss}, {Colazo}, {Dahal}, {Denis},
  {D{\"u}nner}, {Eimer}, {Essinger-Hileman}, {Fluxa}, {Halpern}, {Hilton},
  {Hinshaw}, {Hubmayr}, {Iuliano}, {Karakla}, {McMahon}, {Miller}, {Moseley},
  {Palma}, {Parker}, {Petroff}, {Pradenas}, {Rostem}, {Sagliocca}, {Valle},
  {Watts}, {Wollack}, {Xu}, \& {Zeng}}]{harrington16spie}
{Harrington}, K., {Marriage}, T., {Ali}, A., {et~al.} 2016, in Society of
  Photo-Optical Instrumentation Engineers (SPIE) Conference Series, Vol. 9914,
  Millimeter, Submillimeter, and Far-Infrared Detectors and Instrumentation for
  Astronomy VIII, ed. W.~S. {Holland} \& J.~{Zmuidzinas}, 99141K,
  \dodoi{10.1117/12.2233125}

\bibitem[{{Harrington} {et~al.}(2018){Harrington}, {Eimer}, {Chuss}, {Petroff},
  {Cleary}, {DeGeorge}, {Grunberg}, {Ali}, {Appel}, {Bennett}, {Brewer},
  {Bustos}, {Chan}, {Couto}, {Dahal}, {Denis}, {D{\"u}nner},
  {Essinger-Hileman}, {Fluxa}, {Halpern}, {Hilton}, {Hinshaw}, {Hubmayr},
  {Iuliano}, {Karakla}, {Marriage}, {McMahon}, {Miller}, {Nu{\~n}ez},
  {Padilla}, {Palma}, {Parker}, {Pradenas Marquez}, {Reeves}, {Reintsema},
  {Rostem}, {Augusto Nunes Valle}, {Van Engelhoven}, {Wang}, {Wang}, {Watts},
  {Weiland}, {Wollack}, {Xu}, {Yan}, \& {Zeng}}]{harrington18spie}
{Harrington}, K., {Eimer}, J., {Chuss}, D.~T., {et~al.} 2018, in Society of
  Photo-Optical Instrumentation Engineers (SPIE) Conference Series, Vol. 10708,
  Society of Photo-Optical Instrumentation Engineers (SPIE) Conference Series,
  107082M, \dodoi{10.1117/12.2313614}

\bibitem[{{Hinshaw} {et~al.}(2013){Hinshaw}, {Larson}, {Komatsu}, {Spergel},
  {Bennett}, {Dunkley}, {Nolta}, {Halpern}, {Hill}, {Odegard}, {Page}, {Smith},
  {Weiland}, {Gold}, {Jarosik}, {Kogut}, {Limon}, {Meyer}, {Tucker}, {Wollack},
  \& {Wright}}]{hinshaw13}
{Hinshaw}, G., {Larson}, D., {Komatsu}, E., {et~al.} 2013, \apjs, 208, 19,
  \dodoi{10.1088/0067-0049/208/2/19}

\bibitem[{{Hui} {et~al.}(2016){Hui}, {Ade}, {Ahmed}, {Alexander}, {Amiri},
  {Barkats}, {Benton}, {Bischoff}, {Bock}, {Boenish}, {Bowens-Rubin},
  {Bowens-Rubin}, {Buder}, {Bullock}, {Buza}, {Connors}, {Filippini},
  {Fliescher}, {Grayson}, {Halpern}, {Harrison}, {Hilton}, {Hristov}, {Irwin},
  {Kang}, {Karkare}, {Karpel}, {Kefeli}, {Kernasovskiy}, {Kovac}, {Kuo},
  {Leitch}, {Lueker}, {Megerian}, {Monticue}, {Namikawa}, {Netterfield},
  {Nguyen}, {O'Brient}, {Ogburn}, {Pryke}, {Reintsema}, {Richter}, {Schwarz},
  {Sorensen}, {Sheehy}, {Staniszewski}, {Steinbach}, {Teply}, {Thompson},
  {Tolan}, {Tucker}, {Turner}, {Vieregg}, {Wandui}, {Weber}, {Wiebe},
  {Willmert}, {Wu}, \& {Yoon}}]{hui16}
{Hui}, H., {Ade}, P.~A.~R., {Ahmed}, Z., {et~al.} 2016, in Society of
  Photo-Optical Instrumentation Engineers (SPIE) Conference Series, Vol. 9914,
  Millimeter, Submillimeter, and Far-Infrared Detectors and Instrumentation for
  Astronomy VIII, ed. W.~S. {Holland} \& J.~{Zmuidzinas}, 99140T,
  \dodoi{10.1117/12.2232986}

\bibitem[{{Irwin} \& {Hilton}(2005)}]{TES-chapter}
{Irwin}, K.~D., \& {Hilton}, G.~C. 2005, in Cryogenic Particle Detection, ed.
  C.~{Enss}, Vol.~99, 63, \dodoi{10.1007/10933596\_3}

\bibitem[{{Iuliano}(2020)}]{iulianophdthesis}
{Iuliano}, J. 2020, PhD thesis, Johns Hopkins University

\bibitem[{{Iuliano} {et~al.}(2018){Iuliano}, {Eimer}, {Parker}, {Rhoades},
  {Ali}, {Appel}, {Bennett}, {Brewer}, {Bustos}, {Chuss}, {Cleary}, {Couto},
  {Dahal}, {Denis}, {D{\"u}nner}, {Essinger-Hileman}, {Fluxa}, {Halpern},
  {Harrington}, {Helson}, {Hilton}, {Hinshaw}, {Hubmayr}, {Karakla},
  {Marriage}, {Miller}, {McMahon}, {Nu{\~n}ez}, {Padilla}, {Palma}, {Petroff},
  {Pradenas M{\'a}rquez}, {Reeves}, {Reintsema}, {Rostem}, {Augusto Nunes
  Valle}, {Van Engelhoven}, {Wang}, {Wang}, {Watts}, {Weiland}, {Wollack},
  {Xu}, {Yan}, \& {Zeng}}]{iuliano2018spie}
{Iuliano}, J., {Eimer}, J., {Parker}, L., {et~al.} 2018, in Society of
  Photo-Optical Instrumentation Engineers (SPIE) Conference Series, Vol. 10708,
  Society of Photo-Optical Instrumentation Engineers (SPIE) Conference Series,
  1070828, \dodoi{10.1117/12.2312954}

\bibitem[{Kittel(1996)}]{kittel96}
Kittel, C. 1996, Introduction to Solid State Physics, 7th edn. (New York: John
  Wiley \& Sons)

\bibitem[{{Kuo} {et~al.}(2008){Kuo}, {Bock}, {Bonetti}, {Brevik},
  {Chattopadhyay}, {Day}, {Golwala}, {Kenyon}, {Lange}, {LeDuc}, {Nguyen},
  {Ogburn}, {Orlando}, {Transgrud}, {Turner}, {Wang}, \& {Zmuidzinas}}]{kuo08}
{Kuo}, C.~L., {Bock}, J.~J., {Bonetti}, J.~A., {et~al.} 2008, in Society of
  Photo-Optical Instrumentation Engineers (SPIE) Conference Series, Vol. 7020,
  Millimeter and Submillimeter Detectors and Instrumentation for Astronomy IV,
  ed. W.~D. {Duncan}, W.~S. {Holland}, S.~{Withington}, \& J.~{Zmuidzinas},
  70201I, \dodoi{10.1117/12.788588}

\bibitem[{{Lazear} {et~al.}(2014){Lazear}, {Ade}, {Benford}, {Bennett},
  {Chuss}, {Dotson}, {Eimer}, {Fixsen}, {Halpern}, {Hilton}, {Hinderks},
  {Hinshaw}, {Irwin}, {Jhabvala}, {Johnson}, {Kogut}, {Lowe}, {McMahon},
  {Miller}, {Mirel}, {Moseley}, {Rodriguez}, {Sharp}, {Staguhn}, {Switzer},
  {Tucker}, {Weston}, \& {Wollack}}]{piper14}
{Lazear}, J., {Ade}, P. A.~R., {Benford}, D., {et~al.} 2014, in Society of
  Photo-Optical Instrumentation Engineers (SPIE) Conference Series, Vol. 9153,
  Millimeter, Submillimeter, and Far-Infrared Detectors and Instrumentation for
  Astronomy VII, ed. W.~S. {Holland} \& J.~{Zmuidzinas}, 91531L,
  \dodoi{10.1117/12.2056806}

\bibitem[{{Lee} {et~al.}(2020){Lee}, {Choi}, {G{\'e}nova-Santos}, {Hattori},
  {Hazumi}, {Honda}, {Ikemitsu}, {Ishida}, {Ishitsuka}, {Jo}, {Karatsu},
  {Kiuchi}, {Komine}, {Koyano}, {Kutsuma}, {Mima}, {Minowa}, {Moon}, {Nagai},
  {Nagasaki}, {Naruse}, {Oguri}, {Otani}, {Peel}, {Rebolo},
  {Rubi{\~n}o-Mart{\'\i}n}, {Sekimoto}, {Suzuki}, {Taino}, {Tajima}, {Tomita},
  {Uchida}, {Won}, \& {Yoshida}}]{groundbird20}
{Lee}, K., {Choi}, J., {G{\'e}nova-Santos}, R.~T., {et~al.} 2020, Journal of
  Low Temperature Physics, 200, 384, \dodoi{10.1007/s10909-020-02511-5}

\bibitem[{{LiteBIRD Collaboration} {et~al.}(2022){LiteBIRD Collaboration},
  {Allys}, {Arnold}, {Aumont}, {Aurlien}, {Azzoni}, {Baccigalupi}, {Banday},
  {Banerji}, {Barreiro}, {Bartolo}, {Bautista}, {Beck}, {Beckman},
  {Bersanelli}, {Boulanger}, {Brilenkov}, {Bucher}, {Calabrese}, {Campeti},
  {Carones}, {Casas}, {Catalano}, {Chan}, {Cheung}, {Chinone}, {Clark},
  {Columbro}, {D'Alessandro}, {de Bernardis}, {de Haan}, {de la Hoz}, {De
  Petris}, {Della Torre}, {Diego-Palazuelos}, {Dobbs}, {Dotani}, {Duval},
  {Elleflot}, {Eriksen}, {Errard}, {Essinger-Hileman}, {Finelli}, {Flauger},
  {Franceschet}, {Fuskeland}, {Galloway}, {Ganga}, {Gerbino}, {Gervasi},
  {G{\'e}nova-Santos}, {Ghigna}, {Giardiello}, {Gjerl{\o}w}, {Grain}, {Grupp},
  {Gruppuso}, {Gudmundsson}, {Halverson}, {Hargrave}, {Hasebe}, {Hasegawa},
  {Hazumi}, {Henrot-Versill{\'e}}, {Hensley}, {Hergt}, {Herman}, {Hivon},
  {Hlozek}, {Hornsby}, {Hoshino}, {Hubmayr}, {Ichiki}, {Iida}, {Imada},
  {Ishino}, {Jaehnig}, {Katayama}, {Kato}, {Keskitalo}, {Kisner}, {Kobayashi},
  {Kogut}, {Kohri}, {Komatsu}, {Komatsu}, {Konishi}, {Krachmalnicoff}, {Kuo},
  {Lamagna}, {Lattanzi}, {Lee}, {Leloup}, {Levrier}, {Linder}, {Luzzi},
  {Macias-Perez}, {Maciaszek}, {Maffei}, {Maino}, {Mandelli},
  {Mart{\'\i}nez-Gonz{\'a}lez}, {Masi}, {Massa}, {Matarrese}, {Matsuda},
  {Matsumura}, {Mele}, {Migliaccio}, {Minami}, {Moggi}, {Montgomery},
  {Montier}, {Morgante}, {Mot}, {Nagano}, {Nagasaki}, {Nagata}, {Nakano},
  {Namikawa}, {Nati}, {Natoli}, {Nerval}, {Noviello}, {Odagiri}, {Oguri},
  {Ohsaki}, {Pagano}, {Paiella}, {Paoletti}, {Passerini}, {Patanchon},
  {Piacentini}, {Piat}, {Polenta}, {Poletti}, {Prouv{\'e}}, {Puglisi},
  {Rambaud}, {Raum}, {Realini}, {Reinecke}, {Remazeilles}, {Ritacco}, {Roudil},
  {Rubino-Martin}, {Russell}, {Sakurai}, {Sakurai}, {Sasaki}, {Scott},
  {Sekimoto}, {Shinozaki}, {Shiraishi}, {Shirron}, {Signorelli}, {Spinella},
  {Stever}, {Stompor}, {Sugiyama}, {Sullivan}, {Suzuki}, {Svalheim}, {Switzer},
  {Takaku}, {Takakura}, {Takase}, {Tartari}, {Terao}, {Thermeau}, {Thommesen},
  {Thompson}, {Tomasi}, {Tominaga}, {Tristram}, {Tsuji}, {Tsujimoto}, {Vacher},
  {Vielva}, {Vittorio}, {Wang}, {Watanuki}, {Wehus}, {Weller}, {Westbrook},
  {Wilms}, {Wollack}, {Yumoto}, \& {Zannoni}}]{litebird22}
{LiteBIRD Collaboration}, {Allys}, E., {Arnold}, K., {et~al.} 2022, arXiv
  e-prints, arXiv:2202.02773, \dodoi{10.48550/arXiv.2202.02773}

\bibitem[{May {et~al.}(2024)May, Adler, Austermann, Benton, Bihary, Durkin,
  Duff, Filippini, Fraisse, Gascard, Gibbs, Gourapura, Gudmundsson, Hartley,
  Hubmayr, Jones, Li, Nagy, Okun, Padilla, Romualdez, Tartakovsky, \&
  Vissers}]{may2024}
May, J.~L., Adler, A.~E., Austermann, J.~E., {et~al.} 2024, in Ground-based and
  Airborne Telescopes X, ed. H.~K. Marshall, J.~Spyromilio, \& T.~Usuda, Vol.
  13094, International Society for Optics and Photonics (SPIE), 1309432,
  \dodoi{10.1117/12.3019051}

\bibitem[{Morgan \& Pan(2013)}]{Morgan2013}
Morgan, M.~A., \& Pan, S.-K. 2013, IEEE Transactions on Terahertz Science and
  Technology, 3, 72, \dodoi{10.1109/TTHZ.2012.2235910}

\bibitem[{{Nagler} {et~al.}(2020){Nagler}, {Sadleir}, \&
  {Wollack}}]{proximity1}
{Nagler}, P.~C., {Sadleir}, J.~E., \& {Wollack}, E.~J. 2020, arXiv e-prints,
  arXiv:2012.06543.
\newblock \doarXiv{2012.06543}

\bibitem[{{Niemack} {et~al.}(2008){Niemack}, {Zhao}, {Wollack}, {Thornton},
  {Switzer}, {Swetz}, {Staggs}, {Page}, {Stryzak}, {Moseley}, {Marriage},
  {Limon}, {Lau}, {Klein}, {Kaul}, {Jarosik}, {Irwin}, {Hincks}, {Hilton},
  {Halpern}, {Fowler}, {Fisher}, {D{\"u}nner}, {Doriese}, {Dicker}, {Devlin},
  {Chervenak}, {Burger}, {Battistelli}, {Appel}, {Amiri}, {Allen}, \&
  {Aboobaker}}]{niemack08}
{Niemack}, M.~D., {Zhao}, Y., {Wollack}, E., {et~al.} 2008, Journal of Low
  Temperature Physics, 151, 690, \dodoi{10.1007/s10909-008-9729-2}

\bibitem[{{N{\'u}{\~n}ez} {et~al.}(2022){N{\'u}{\~n}ez}, {Appel}, {Bruno},
  {Datta}, {Ali}, {Bennett}, {Dahal}, {Denes Couto}, {Denis}, {Eimer},
  {Espinoza}, {Essinger-Hileman}, {Helson}, {Iuliano}, {Marriage}, {Morales
  Per{\'e}z}, {Nunes Valle}, {Petroff}, {Rostem}, {Shi}, {Watts}, {Wollack}, \&
  {Xu}}]{nunez22spie}
{N{\'u}{\~n}ez}, C., {Appel}, J.~W., {Bruno}, S.~M., {et~al.} 2022, in Society
  of Photo-Optical Instrumentation Engineers (SPIE) Conference Series, Vol.
  12190, Millimeter, Submillimeter, and Far-Infrared Detectors and
  Instrumentation for Astronomy XI, ed. J.~{Zmuidzinas} \& J.-R. {Gao},
  121901J, \dodoi{10.1117/12.2630081}

\bibitem[{{N{\'u}{\~n}ez} {et~al.}(2023){N{\'u}{\~n}ez}, {Appel}, {Brewer},
  {Bruno}, {Datta}, {Bennett}, {Bustos}, {Chuss}, {Dahal}, {Denis}, {Eimer},
  {Essinger-Hileman}, {Helson}, {Marriage}, {P{\'e}rez}, {Padilla}, {Petroff},
  {Rostem}, {Watts}, {Wollack}, \& {Xu}}]{nunez23asc}
{N{\'u}{\~n}ez}, C., {Appel}, J.~W., {Brewer}, M.~K., {et~al.} 2023, IEEE
  Transactions on Applied Superconductivity, 33, 3262497,
  \dodoi{10.1109/TASC.2023.3262497}

\bibitem[{{Pan} {et~al.}(2019){Pan}, {Liu}, {Basu Thakur}, {Benson}, {Fixsen},
  {Goksu}, {Rath}, \& {Meyer}}]{Pan2019}
{Pan}, Z., {Liu}, M., {Basu Thakur}, R., {et~al.} 2019, \ao, 58, 6257,
  \dodoi{10.1364/AO.58.006257}

\bibitem[{{Planck Collaboration} {et~al.}(2016){Planck Collaboration}, {Ade},
  {Aghanim}, {Ashdown}, {Aumont}, {Baccigalupi}, {Banday}, {Barreiro},
  {Bartolo}, {Battaglia}, \& et~al.}]{planck15V}
{Planck Collaboration}, {Ade}, P.~A.~R., {Aghanim}, N., {et~al.} 2016, \aap,
  594, A5, \dodoi{10.1051/0004-6361/201526632}

\bibitem[{{Planck Collaboration} {et~al.}(2020){Planck Collaboration},
  {Aghanim}, {Akrami}, {Ashdown}, {Aumont}, {Baccigalupi}, {Ballardini},
  {Banday}, {Barreiro}, {Bartolo}, {Basak}, {Battye}, {Benabed}, {Bernard},
  {Bersanelli}, {Bielewicz}, {Bock}, {Bond}, {Borrill}, {Bouchet}, {Boulanger},
  {Bucher}, {Burigana}, {Butler}, {Calabrese}, {Cardoso}, {Carron},
  {Challinor}, {Chiang}, {Chluba}, {Colombo}, {Combet}, {Contreras}, {Crill},
  {Cuttaia}, {de Bernardis}, {de Zotti}, {Delabrouille}, {Delouis}, {Di
  Valentino}, {Diego}, {Dor{\'e}}, {Douspis}, {Ducout}, {Dupac}, {Dusini},
  {Efstathiou}, {Elsner}, {En{\ss}lin}, {Eriksen}, {Fantaye}, {Farhang},
  {Fergusson}, {Fernandez-Cobos}, {Finelli}, {Forastieri}, {Frailis},
  {Fraisse}, {Franceschi}, {Frolov}, {Galeotta}, {Galli}, {Ganga},
  {G{\'e}nova-Santos}, {Gerbino}, {Ghosh}, {Gonz{\'a}lez-Nuevo}, {G{\'o}rski},
  {Gratton}, {Gruppuso}, {Gudmundsson}, {Hamann}, {Handley}, {Hansen},
  {Herranz}, {Hildebrandt}, {Hivon}, {Huang}, {Jaffe}, {Jones}, {Karakci},
  {Keih{\"a}nen}, {Keskitalo}, {Kiiveri}, {Kim}, {Kisner}, {Knox},
  {Krachmalnicoff}, {Kunz}, {Kurki-Suonio}, {Lagache}, {Lamarre}, {Lasenby},
  {Lattanzi}, {Lawrence}, {Le Jeune}, {Lemos}, {Lesgourgues}, {Levrier},
  {Lewis}, {Liguori}, {Lilje}, {Lilley}, {Lindholm}, {L{\'o}pez-Caniego},
  {Lubin}, {Ma}, {Mac{\'\i}as-P{\'e}rez}, {Maggio}, {Maino}, {Mandolesi},
  {Mangilli}, {Marcos-Caballero}, {Maris}, {Martin}, {Martinelli},
  {Mart{\'\i}nez-Gonz{\'a}lez}, {Matarrese}, {Mauri}, {McEwen}, {Meinhold},
  {Melchiorri}, {Mennella}, {Migliaccio}, {Millea}, {Mitra},
  {Miville-Desch{\^e}nes}, {Molinari}, {Montier}, {Morgante}, {Moss}, {Natoli},
  {N{\o}rgaard-Nielsen}, {Pagano}, {Paoletti}, {Partridge}, {Patanchon},
  {Peiris}, {Perrotta}, {Pettorino}, {Piacentini}, {Polastri}, {Polenta},
  {Puget}, {Rachen}, {Reinecke}, {Remazeilles}, {Renzi}, {Rocha}, {Rosset},
  {Roudier}, {Rubi{\~n}o-Mart{\'\i}n}, {Ruiz-Granados}, {Salvati}, {Sandri},
  {Savelainen}, {Scott}, {Shellard}, {Sirignano}, {Sirri}, {Spencer},
  {Sunyaev}, {Suur-Uski}, {Tauber}, {Tavagnacco}, {Tenti}, {Toffolatti},
  {Tomasi}, {Trombetti}, {Valenziano}, {Valiviita}, {Van Tent}, {Vibert},
  {Vielva}, {Villa}, {Vittorio}, {Wandelt}, {Wehus}, {White}, {White},
  {Zacchei}, \& {Zonca}}]{planck18VI}
{Planck Collaboration}, {Aghanim}, N., {Akrami}, Y., {et~al.} 2020, \aap, 641,
  A6, \dodoi{10.1051/0004-6361/201833910}

\bibitem[{{Posada} {et~al.}(2016){Posada}, {Ade}, {Anderson}, {Avva}, {Ahmed},
  {Arnold}, {Austermann}, {Bender}, {Benson}, {Bleem}, {Byrum}, {Carlstrom},
  {Carter}, {Chang}, {Cho}, {Cukierman}, {Czaplewski}, {Ding}, {Divan}, {de
  Haan}, {Dobbs}, {Dutcher}, {Everett}, {Gannon}, {Guyser}, {Halverson},
  {Harrington}, {Hattori}, {Henning}, {Hilton}, {Holzapfel}, {Huang}, {Irwin},
  {Jeong}, {Khaire}, {Korman}, {Kubik}, {Kuo}, {Lee}, {Leitch}, {Lendinez
  Escudero}, {Meyer}, {Miller}, {Montgomery}, {Nadolski}, {Natoli}, {Nguyen},
  {Novosad}, {Padin}, {Pan}, {Pearson}, {Rahlin}, {Reichardt}, {Ruhl},
  {Saliwanchik}, {Shirley}, {Sayre}, {Shariff}, {Shirokoff}, {Stan}, {Stark},
  {Sobrin}, {Story}, {Suzuki}, {Tang}, {Thakur}, {Thompson}, {Tucker},
  {Vanderlinde}, {Vieira}, {Wang}, {Whitehorn}, {Yefremenko}, \&
  {Yoon}}]{posada16}
{Posada}, C.~M., {Ade}, P. A.~R., {Anderson}, A.~J., {et~al.} 2016, in Society
  of Photo-Optical Instrumentation Engineers (SPIE) Conference Series, Vol.
  9914, Millimeter, Submillimeter, and Far-Infrared Detectors and
  Instrumentation for Astronomy VIII, ed. W.~S. {Holland} \& J.~{Zmuidzinas},
  991417, \dodoi{10.1117/12.2232912}

\bibitem[{{Reintsema} {et~al.}(2003){Reintsema}, {Beyer}, {Nam}, {Deiker},
  {Hilton}, {Irwin}, {Martinis}, {Ullom}, {Vale}, \& {Macintosh}}]{reintsema03}
{Reintsema}, C.~D., {Beyer}, J., {Nam}, S.~W., {et~al.} 2003, Review of
  Scientific Instruments, 74, 4500, \dodoi{10.1063/1.1605259}

\bibitem[{{Rostem} {et~al.}(2012){Rostem}, {Bennett}, {Chuss}, {Costen},
  {Crowe}, {Denis}, {Eimer}, {Lourie}, {Essinger-Hileman}, {Marriage},
  {Moseley}, {Stevenson}, {Towner}, {Voellmer}, {Wollack}, \&
  {Zeng}}]{rostem12spie}
{Rostem}, K., {Bennett}, C.~L., {Chuss}, D.~T., {et~al.} 2012, in Society of
  Photo-Optical Instrumentation Engineers (SPIE) Conference Series, Vol. 8452,
  Millimeter, Submillimeter, and Far-Infrared Detectors and Instrumentation for
  Astronomy VI, ed. W.~S. {Holland} \& J.~{Zmuidzinas}, 84521N,
  \dodoi{10.1117/12.927056}

\bibitem[{{Rostem} {et~al.}(2014){Rostem}, {Chuss}, {Colazo}, {Crowe}, {Denis},
  {Lourie}, {Moseley}, {Stevenson}, \& {Wollack}}]{rostem-ballistic}
{Rostem}, K., {Chuss}, D.~T., {Colazo}, F.~A., {et~al.} 2014, Journal of
  Applied Physics, 115, 124508, \dodoi{10.1063/1.4869737}

\bibitem[{{Rostem} {et~al.}(2016){Rostem}, {Ali}, {Appel}, {Bennett}, {Brown},
  {Chang}, {Chuss}, {Colazo}, {Costen}, {Denis}, {Essinger-Hileman}, {Hu},
  {Marriage}, {Moseley}, {Stevenson}, {U-Yen}, {Wollack}, \&
  {Xu}}]{rostem16spie}
{Rostem}, K., {Ali}, A., {Appel}, J.~W., {et~al.} 2016, in Society of
  Photo-Optical Instrumentation Engineers (SPIE) Conference Series, Vol. 9914,
  Millimeter, Submillimeter, and Far-Infrared Detectors and Instrumentation for
  Astronomy VIII, ed. W.~S. {Holland} \& J.~{Zmuidzinas}, 99140D,
  \dodoi{10.1117/12.2234308}

\bibitem[{{Rubi{\~n}o-Mart{\'\i}n} {et~al.}(2023){Rubi{\~n}o-Mart{\'\i}n},
  {Guidi}, {G{\'e}nova-Santos}, {Harper}, {Herranz}, {Hoyland}, {Lasenby},
  {Poidevin}, {Rebolo}, {Ruiz-Granados}, {Vansyngel}, {Vielva}, {Watson},
  {Artal}, {Ashdown}, {Barreiro}, {Bilbao-Ahedo}, {Casas}, {Casaponsa},
  {Cepeda-Arroita}, {de la Hoz}, {Dickinson}, {Fern{\'a}ndez-Cobos},
  {Fern{\'a}ndez-Torreiro}, {Gonz{\'a}lez-Gonz{\'a}lez},
  {Hern{\'a}ndez-Monteagudo}, {L{\'o}pez-Caniego}, {L{\'o}pez-Caraballo},
  {Mart{\'\i}nez-Gonz{\'a}lez}, {Peel}, {Pel{\'a}ez-Santos}, {Perrott},
  {Piccirillo}, {Razavi-Ghods}, {Scott}, {Titterington}, {Tramonte}, \&
  {Vignaga}}]{quijote23}
{Rubi{\~n}o-Mart{\'\i}n}, J.~A., {Guidi}, F., {G{\'e}nova-Santos}, R.~T.,
  {et~al.} 2023, \mnras, 519, 3383, \dodoi{10.1093/mnras/stac3439}

\bibitem[{{Sadleir} {et~al.}(2010){Sadleir}, {Smith}, {Bandler}, {Chervenak},
  \& {Clem}}]{proximity2}
{Sadleir}, J.~E., {Smith}, S.~J., {Bandler}, S.~R., {Chervenak}, J.~A., \&
  {Clem}, J.~R. 2010, \prl, 104, 047003, \dodoi{10.1103/PhysRevLett.104.047003}

\bibitem[{Shirokoff {et~al.}(2009)Shirokoff, Benson, Bleem, Chang, Cho, Crites,
  Dobbs, Holzapfel, Lanting, Lee, Lueker, Mehl, Plagge, Spieler, \&
  Vieira}]{shirokoff10}
Shirokoff, E., Benson, B.~A., Bleem, L.~E., {et~al.} 2009, IEEE Transactions on
  Applied Superconductivity, 19, 517, \dodoi{10.1109/TASC.2009.2018229}

\bibitem[{{Simons Observatory Collaboration} {et~al.}(2019){Simons Observatory
  Collaboration}, {Ade}, {Aguirre}, {Ahmed}, {Aiola}, {Ali}, {Alonso},
  {Alvarez}, {Arnold}, {Ashton}, {Austermann}, {Awan}, {Baccigalupi},
  {Baildon}, {Barron}, {Battaglia}, {Battye}, {Baxter}, {Bazarko}, {Beall},
  {Bean}, {Beck}, {Beckman}, {Beringue}, {Bianchini}, {Boada}, {Boettger},
  {Bond}, {Borrill}, {Brown}, {Bruno}, {Bryan}, {Calabrese}, {Calafut},
  {Calisse}, {Carron}, {Challinor}, {Chesmore}, {Chinone}, {Chluba}, {Cho},
  {Choi}, {Coppi}, {Cothard}, {Coughlin}, {Crichton}, {Crowley}, {Crowley},
  {Cukierman}, {D'Ewart}, {D{\"u}nner}, {de Haan}, {Devlin}, {Dicker},
  {Didier}, {Dobbs}, {Dober}, {Duell}, {Duff}, {Duivenvoorden}, {Dunkley},
  {Dusatko}, {Errard}, {Fabbian}, {Feeney}, {Ferraro}, {Flux{\`a}}, {Freese},
  {Frisch}, {Frolov}, {Fuller}, {Fuzia}, {Galitzki}, {Gallardo}, {Tomas Galvez
  Ghersi}, {Gao}, {Gawiser}, {Gerbino}, {Gluscevic}, {Goeckner-Wald}, {Golec},
  {Gordon}, {Gralla}, {Green}, {Grigorian}, {Groh}, {Groppi}, {Guan},
  {Gudmundsson}, {Han}, {Hargrave}, {Hasegawa}, {Hasselfield}, {Hattori},
  {Haynes}, {Hazumi}, {He}, {Healy}, {Henderson}, {Hervias-Caimapo}, {Hill},
  {Hill}, {Hilton}, {Hilton}, {Hincks}, {Hinshaw}, {Hlo{\v{z}}ek}, {Ho}, {Ho},
  {Howe}, {Huang}, {Hubmayr}, {Huffenberger}, {Hughes}, {Ijjas}, {Ikape},
  {Irwin}, {Jaffe}, {Jain}, {Jeong}, {Kaneko}, {Karpel}, {Katayama}, {Keating},
  {Kernasovskiy}, {Keskitalo}, {Kisner}, {Kiuchi}, {Klein}, {Knowles},
  {Koopman}, {Kosowsky}, {Krachmalnicoff}, {Kuenstner}, {Kuo}, {Kusaka},
  {Lashner}, {Lee}, {Lee}, {Leon}, {Leung}, {Lewis}, {Li}, {Li}, {Limon},
  {Linder}, {Lopez-Caraballo}, {Louis}, {Lowry}, {Lungu}, {Madhavacheril},
  {Mak}, {Maldonado}, {Mani}, {Mates}, {Matsuda}, {Maurin}, {Mauskopf}, {May},
  {McCallum}, {McKenney}, {McMahon}, {Meerburg}, {Meyers}, {Miller},
  {Mirmelstein}, {Moodley}, {Munchmeyer}, {Munson}, {Naess}, {Nati},
  {Navaroli}, {Newburgh}, {Nguyen}, {Niemack}, {Nishino}, {Orlowski-Scherer},
  {Page}, {Partridge}, {Peloton}, {Perrotta}, {Piccirillo}, {Pisano},
  {Poletti}, {Puddu}, {Puglisi}, {Raum}, {Reichardt}, {Remazeilles},
  {Rephaeli}, {Riechers}, {Rojas}, {Roy}, {Sadeh}, {Sakurai}, {Salatino},
  {Sathyanarayana Rao}, {Schaan}, {Schmittfull}, {Sehgal}, {Seibert}, {Seljak},
  {Sherwin}, {Shimon}, {Sierra}, {Sievers}, {Sikhosana}, {Silva-Feaver},
  {Simon}, {Sinclair}, {Siritanasak}, {Smith}, {Smith}, {Spergel}, {Staggs},
  {Stein}, {Stevens}, {Stompor}, {Suzuki}, {Tajima}, {Takakura}, {Teply},
  {Thomas}, {Thorne}, {Thornton}, {Trac}, {Tsai}, {Tucker}, {Ullom},
  {Vagnozzi}, {van Engelen}, {Van Lanen}, {Van Winkle}, {Vavagiakis},
  {Verg{\`e}s}, {Vissers}, {Wagoner}, {Walker}, {Ward}, {Westbrook},
  {Whitehorn}, {Williams}, {Williams}, {Wollack}, {Xu}, {Yu}, {Yu}, {Zago},
  {Zhang}, \& {Zhu}}]{simons19whitepaper}
{Simons Observatory Collaboration}, {Ade}, P., {Aguirre}, J., {et~al.} 2019,
  \jcap, 2019, 056, \dodoi{10.1088/1475-7516/2019/02/056}

\bibitem[{{Staguhn} {et~al.}(2004){Staguhn}, {Benford}, {Chervenak}, {Moseley},
  {Allen}, {Stevenson}, \& {Hsieh}}]{Staguhn2004}
{Staguhn}, J.~G., {Benford}, D.~J., {Chervenak}, J.~A., {et~al.} 2004, in
  Society of Photo-Optical Instrumentation Engineers (SPIE) Conference Series,
  Vol. 5498, Z-Spec: a broadband millimeter-wave grating spectrometer: design,
  construction, and first cryogenic measurements, ed. C.~M. {Bradford},
  P.~A.~R. {Ade}, J.~E. {Aguirre}, J.~J. {Bock}, M.~{Dragovan}, L.~{Duband},
  L.~{Earle}, J.~{Glenn}, H.~{Matsuhara}, B.~J. {Naylor}, H.~T. {Nguyen},
  M.~{Yun}, \& J.~{Zmuidzinas}, 390--395, \dodoi{10.1117/12.552102}

\bibitem[{{Suzuki} {et~al.}(2012){Suzuki}, {Arnold}, {Edwards}, {Engargiola},
  {Ghribi}, {Holzapfel}, {Lee}, {Meng}, {Myers}, {O'Brient}, {Quealy},
  {Rebeiz}, \& {Richards}}]{suzuki12}
{Suzuki}, A., {Arnold}, K., {Edwards}, J., {et~al.} 2012, Journal of Low
  Temperature Physics, 167, 852, \dodoi{10.1007/s10909-012-0602-y}

\bibitem[{U-yen \& Wollack(2008)}]{U-yen-Filters}
U-yen, K., \& Wollack, E.~J. 2008, in 2008 38th European Microwave Conference,
  642--645, \dodoi{10.1109/EUMC.2008.4751534}

\bibitem[{U-yen {et~al.}(2009)U-yen, Wollack, Moseley, Stevenson, Hsieh, \&
  Cao}]{Uyen2009}
U-yen, K., Wollack, E.~J., Moseley, S.~H., {et~al.} 2009, in 2009 IEEE MTT-S
  International Microwave Symposium Digest, 1029--1032,
  \dodoi{10.1109/MWSYM.2009.5165875}

\bibitem[{{U-Yen} {et~al.}(2008){U-Yen}, {Wollack}, {Papapolymerou}, \&
  {Laskar}}]{U-Yen-magicT}
{U-Yen}, K., {Wollack}, E.~J., {Papapolymerou}, J., \& {Laskar}, J. 2008, IEEE
  Transactions on Microwave Theory Techniques, 56, 172,
  \dodoi{10.1109/TMTT.2007.912213}

\bibitem[{{Ullom} {et~al.}(2004){Ullom}, {Doriese}, {Hilton}, {Beall},
  {Deiker}, {Duncan}, {Ferreira}, {Irwin}, {Reintsema}, \&
  {Vale}}]{Ullom-zebra}
{Ullom}, J.~N., {Doriese}, W.~B., {Hilton}, G.~C., {et~al.} 2004, Applied
  Physics Letters, 84, 4206, \dodoi{10.1063/1.1753058}

\bibitem[{{Watts} {et~al.}(2015){Watts}, {Larson}, {Marriage}, {Abitbol},
  {Appel}, {Bennett}, {Chuss}, {Eimer}, {Essinger-Hileman}, {Miller}, {Rostem},
  \& {Wollack}}]{watts15}
{Watts}, D.~J., {Larson}, D., {Marriage}, T.~A., {et~al.} 2015, \apj, 814, 103,
  \dodoi{10.1088/0004-637X/814/2/103}

\bibitem[{{Watts} {et~al.}(2018){Watts}, {Wang}, {Ali}, {Appel}, {Bennett},
  {Chuss}, {Dahal}, {Eimer}, {Essinger-Hileman}, {Harrington}, {Hinshaw},
  {Iuliano}, {Marriage}, {Miller}, {Padilla}, {Parker}, {Petroff}, {Rostem},
  {Wollack}, \& {Xu}}]{watts18}
{Watts}, D.~J., {Wang}, B., {Ali}, A., {et~al.} 2018, \apj, 863, 121,
  \dodoi{10.3847/1538-4357/aad283}

\bibitem[{{Wei}(2012)}]{FTS}
{Wei}, T. 2012, Bachelor's thesis, Johns Hopkins University, Maryland

\bibitem[{{Zeng} {et~al.}(2010){Zeng}, {Bennett}, {Chuss}, \&
  {Wollack}}]{zeng10spie}
{Zeng}, L., {Bennett}, C.~L., {Chuss}, D.~T., \& {Wollack}, E.~J. 2010, in
  Society of Photo-Optical Instrumentation Engineers (SPIE) Conference Series,
  Vol. 7741, Millimeter, Submillimeter, and Far-Infrared Detectors and
  Instrumentation for Astronomy V, ed. W.~S. {Holland} \& J.~{Zmuidzinas},
  774129, \dodoi{10.1117/12.857679}

\end{thebibliography}
\bibliographystyle{aasjournal}

\end{document}